\documentclass[a4paper,11pt]{article}

\usepackage{pos}
\usepackage{enumitem}
\usepackage{lineno}
\usepackage{lipsum} %package to generate placeholder text in the following

\title{Highlights from the IceCube Neutrino Observatory}

\ShortTitle{IceCube Highlights}

\author{The IceCube Collaboration \\{\normalsize \normalfont(a complete list of authors can be found at the end of the proceedings)}\\}

\emailAdd{alexander.kappes@uni-muenster.de}

\abstract{
The IceCube neutrino observatory has been successfully operating in its full configuration for almost 15 years and is characterized by a remarkably high stability and up time. During this time, it has made many groundbreaking observations, such as the first detection of a high-energy diffuse cosmic neutrino flux or, more recently, the identification of the AGN NGC~1068 as a steady source of high-energy neutrino emission and the observation of neutrinos from the Milky Way. In this talk, new developments in these areas will be discussed and further highlights presented. The second part then looks at the ongoing developments at the South Pole with IceCube Upgrade and IceCube-Gen2 and discusses their potential for advancing neutrino and astroparticle physics.

\vspace{4mm}

{\bfseries Corresponding authors:}
A.~Kappes$^{1*}$\\
{$^{1}$ \itshape Institute for Nuclear Physics, University Münster}\\[4mm]
$^*$ Presenter
}

\ConferenceLogo{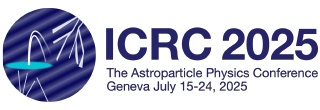}

\FullConference{39th International Cosmic Ray Conference (ICRC2025)\\
 15–24 July 2025\\
Geneva, Switzerland\\}

\begin{document}

\maketitle

\section{Introduction}\label{sec:intro}
One of the central open questions in astroparticle physics concerns the origin and mechanisms of the sources of high-energy cosmic rays, which accelerate particles to energies up to ten million times greater than those achieved by terrestrial accelerators. Over the past two decades, it has become increasingly evident that addressing this problem requires exploiting every available piece of information. This necessity has fostered the emergence of a highly interconnected community that has established infrastructure enabling real-time global data exchange across diverse experiments (e.g.\ \cite{Mancina:2025icrc,Raab:2025icrc}). Since the first observation of a diffuse high-energy astrophysical neutrino flux in 2013 \cite{IceCube:2013low}, these elusive particles have become a key probe in the multimessenger investigation of the high-energy universe. In 2017, the discovery was followed by the first convincing identification of a high-energy neutrino source, the blazar TXS~0506+056, which was identified in a multimessenger campaign with Fermi and other instruments \cite{IceCube:2018dnn,IceCube:2018cha}. The first identification of a neutrino source only with neutrinos occurred in 2022 where IceCube reported the observation of about 80 neutrinos from the active galaxy NGC~1068 \cite{IceCube:2022der}, followed by the first detection of neutrino emission from the Milky Way in 2023 \cite{IceCube:2023ame}. 

\section{The IceCube Neutrino Observatory}\label{sec:icecube}
The IceCube Neutrino Observatory \cite{Aartsen:2016nxy} is currently the largest neutrino telescope worldwide. It instruments $1\,\mathrm{km}^3$ of clear ice in depth between $1450\,\mathrm{m}$ and $2450\,\mathrm{m}$ below the South Pole with 5160 Digital Optical Modules (DOMs) which are attached to 86 vertical strings. A DOM contains a large 10-inch photomultiplier that points downwards together with electronics for the in-situ digitization of signals. They detect the faint Cherenkov light that is emitted from high-energy charged particles which are generated in neutrino interactions in the vicinity of the detector. On their way to the DOMs, Cherenkov photons undergo scattering and absorption in the ice, whose optical properties exhibit strong depth dependence. In the clearest parts of the detector below $2100\,\mathrm{m}$, the absorption length is about $200 \,\mathrm{m}$ and the scattering length $50 \,\mathrm{m}$ at $400\,\mathrm{nm}$ \cite{IceCube:2013llx,Chirkin:2025icrc}. 

Neutrino interactions leave two major signatures in the detector: in case of muon neutrinos undergoing a charged current interaction, a high-energy muon is generated which can travel distances up to several kilometers in the ice and leaves an elongated track-like pattern of light in the detector. On the other hand, charged current electron and tau neutrino as well as all neutral current interactions produce hadronic and/or electromagnetic showers in the ice which, due to the short scattering length of optical light in ice, leads to an almost spherical pattern in the detector. A somewhat special case is that of charged-current tau interactions. The produced tau lepton is extremely short-lived and therefore propagates only a short distance in the ice before decaying, in more than 80\% of cases giving rise to a secondary shower. Only at PeV and higher energies, the showers at the two vertices are significantly far apart that they can be separated visually. However, at lower energies information on such a ``double bang'' pattern can be extracted from the recorded waveforms in the DOMs \cite{IceCube:2024nhk}. 

The event reconstruction involves the determination of the direction of the original neutrino and its energy together with the uncertainties on these quantities. Due to the elongated track-like pattern in the detector, the best pointing precision is achieved with muon neutrinos which undergo a charged current interaction. For these events, angular resolutions of $0.5^\circ$ are achieved above $10\,\mathrm{TeV}$ \cite{IceCube:2021oqo}. In case that the neutrino interaction happens outside the detector, an unknown fraction of the original neutrino energy is transferred into the invisible hadronic shower at the interaction vertex. In addition, the muon loses an unknown fraction of energy on its way to the detector resulting in an energy uncertainty of about a factor 2. On the other hand, the energy of a shower inside the detector can be reconstructed to better than 10\% for energies larger than $10\,\mathrm{TeV}$ \cite{IceCube:2013dkx,Abbasi:2021ryj}. In case of electron neutrino charged current interactions, this is also the energy resolution for the neutrino, whereas in neutral current events some fraction is carried away by the neutrino. The directional reconstruction for contained spherical light pattern is better than $5^\circ$ for energies above $50\,\mathrm{TeV}$ \cite{IceCube:2024csv}. 

The in-ice detector is complemented at the surface by 162 frozen-water tanks with two DOMs each located at the string positions. This detector, named IceTop, detects the shower footprint of cosmic-ray induced air showers including GeV muons. In combination with the in-ice detector which is reached only by TeV muons, the IceCube Neutrino Observatory provides unique capabilities to measure the cosmic-ray composition around the knee. Both the in-ice detector and IceTop allow for measurements of anisotropies in the cosmic-ray flux across different regions of the sky \cite{IceCube:2024pnx}. In addition, IceTop can serve as an air-shower veto for the in-ice detector.   

\section{Diffuse neutrino fluxes}\label{sec:diffuse}

\begin{figure}[t!]
\centering
\includegraphics[width=0.8\linewidth]{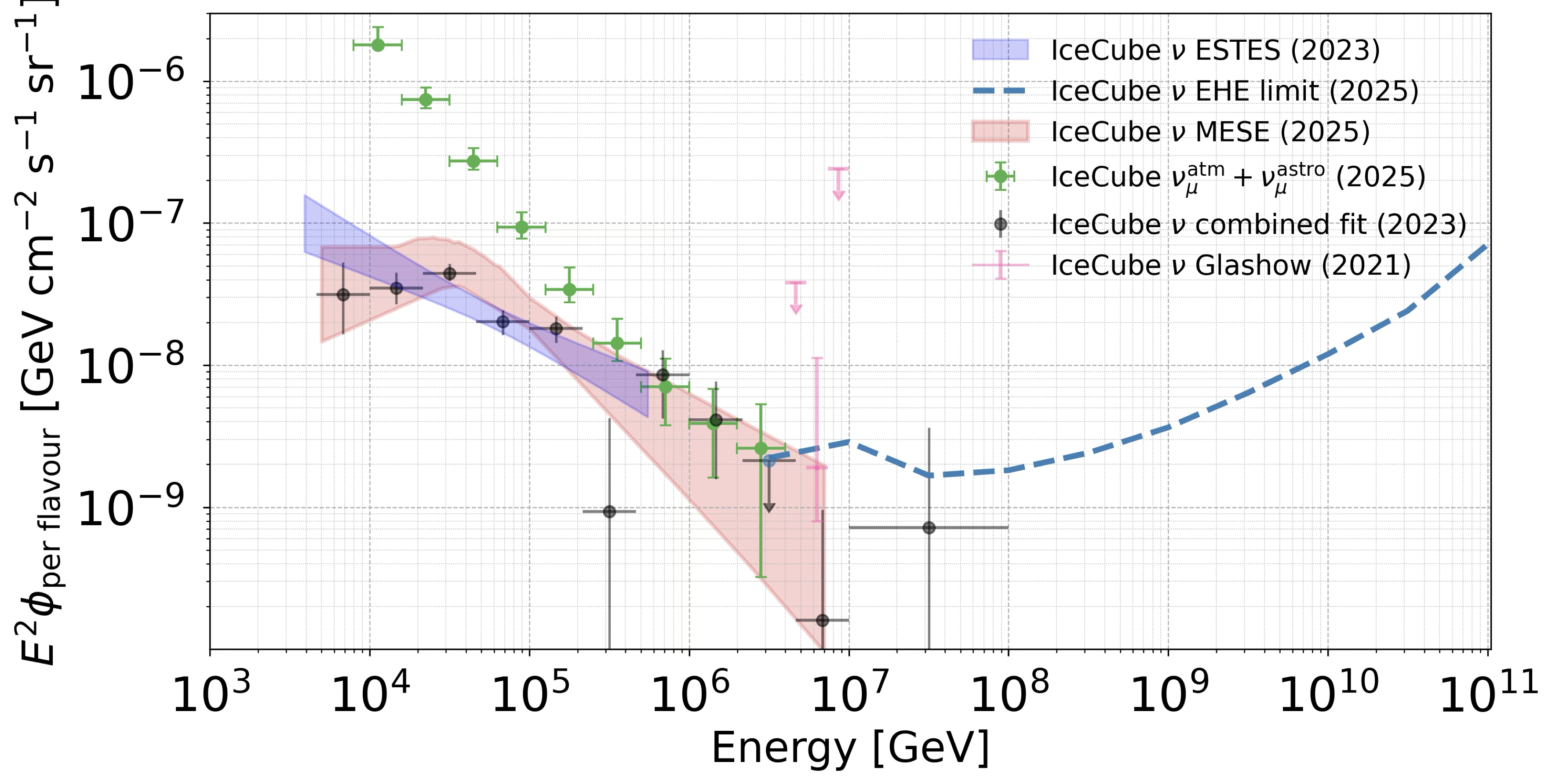}
\caption{Measured diffuse neutrino spectrum per neutrino flavor as a function of energy. The different components shown are: Unfolded total muon neutrino flux (green circles \cite{Rootselar:2025icrc}); Measured cosmic neutrino flux (black circles \cite{IceCube:2025tgp}); 90\% U.L.\ on EHE neutrino above $2\,\mathrm{PeV}$ (\cite{IceCubeCollaborationSS:2025jbi}). Also shown are the 68\% uncertainty bands for two analyses with different event selections (ESTES: blue \cite{IceCube:2024fxo}; MESE: red \cite{IceCube:2025tgp}) and the measured flux at the Glashow resonance \cite{IceCube:2021rpz}.}\label{fig:diffuse_spectrum}
\end{figure}

A large number of neutrinos reaches the IceCube detector continuously from all directions (Fig.~\ref{fig:diffuse_spectrum}). The vast majority of these neutrinos are generated in the atmosphere in air showers initiated by high-energy cosmic rays. One can separate atmospheric neutrinos into two classes. The so called ``conventional'' atmospheric neutrino flux originates from the decay of pions and kaons and follows a power-law spectrum with index of about $3.7$ \cite{Gaisser:2019efm}. It dominates the atmospheric flux at all but the highest energies. The second class is the so called ``prompt'' neutrino flux. It originates from the decay of charmed QCD bound states. Compared to pions and kaons, these states have a much shorter lifetime and therefore most of them decay producing neutrinos instead of getting ``lost'' due to an interaction with nuclei in the atmosphere. This leads to a neutrino flux characterized by a harder power-law index of approximately 2.7 \cite{Gaisser:2019efm}, implying that it should become dominant over the conventional neutrino flux at high energies. However, up to now, no indication of such a prompt neutrino flux has been found in IceCube \cite{Abbasi:2021qfz}. 

\subsection{Astrophysical diffuse neutrino fluxes}\label{sec:astro_diffuse}

\begin{figure}[t!]
\centering
\includegraphics[width=0.45\linewidth]{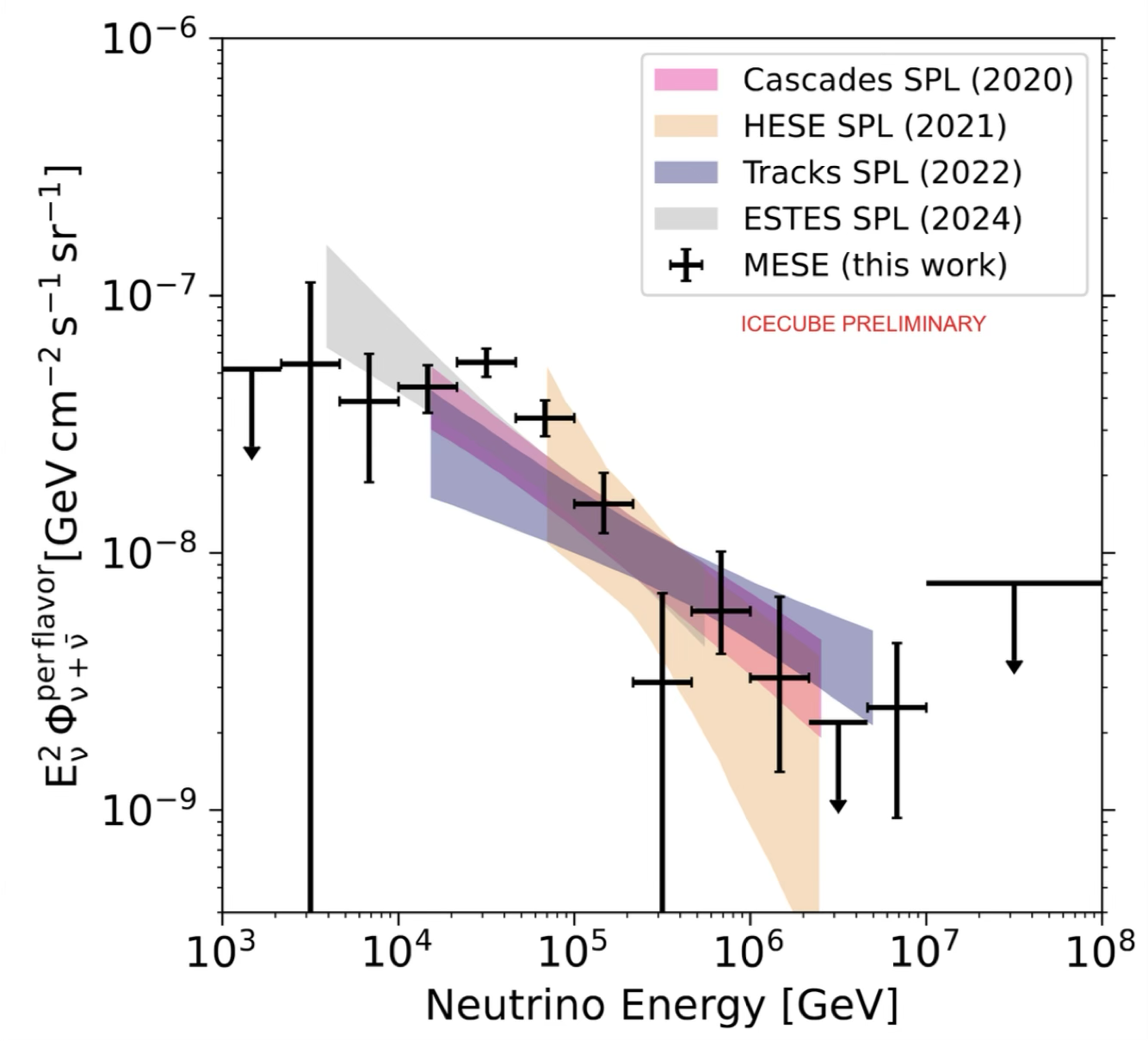}
\hfill
\includegraphics[width=0.45\linewidth]{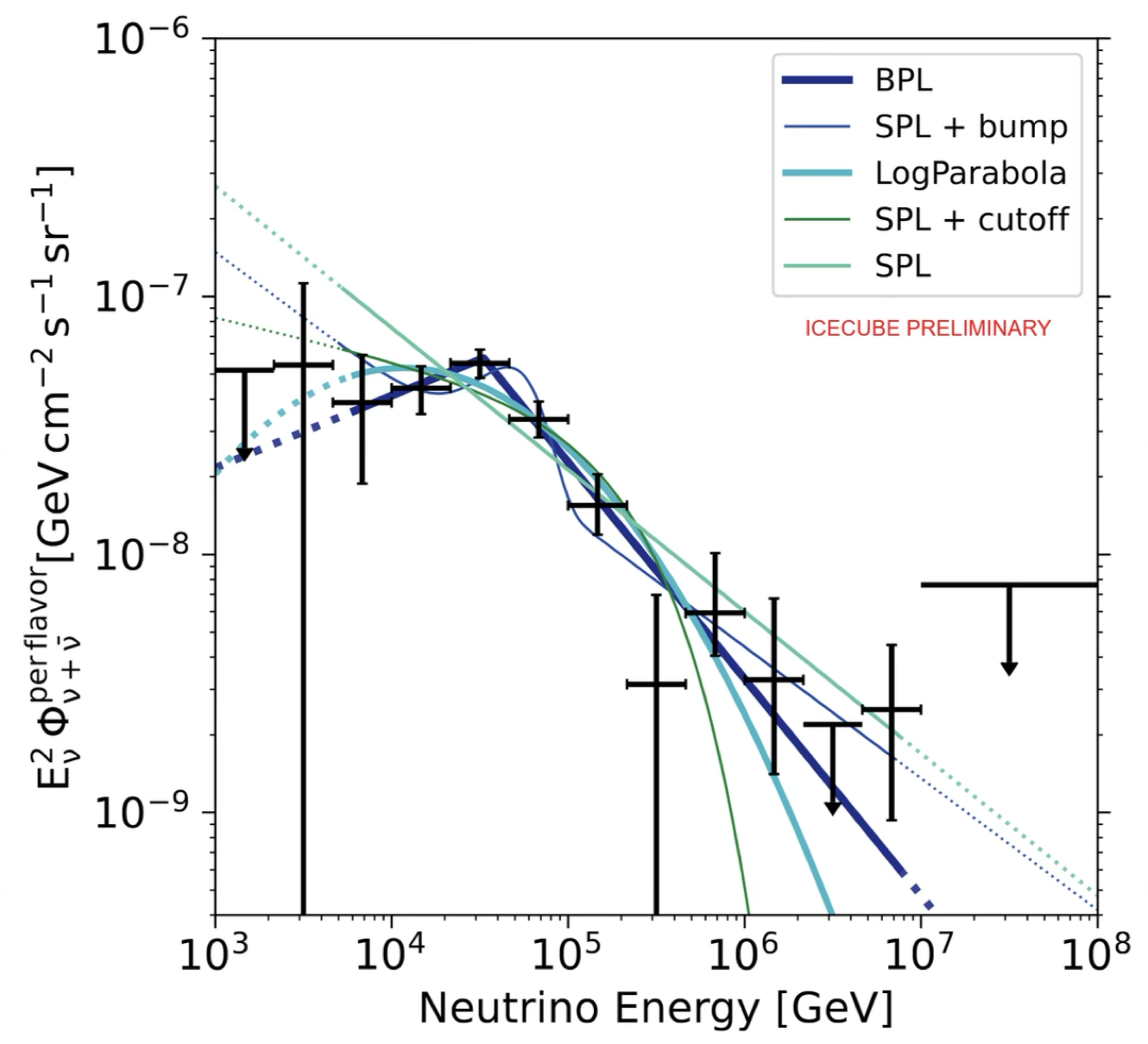}
\caption{\emph{Left:} Measured diffuse astrophysical muon neutrino spectrum together with previous best fit results of a simple power law for different event samples (for details see \cite{Basu:2025icrc}). \emph{Right:} Fit of different spectrum hypotheses: Broken Power-law (BPL), Simple Power-Law (SPL) with bump, Log-parabola, SPL with cutoff and SPL. Both plots taken from \cite{Basu:2025icrc}.} \label{fig:diffuse_break}
\end{figure}

Owing to its expected harder spectrum, the astrophysical diffuse neutrino flux—presumably produced in the most powerful astrophysical sources, such as Active Galactic Nuclei (AGNs), Gamma-Ray Bursts, and supernova remnants—becomes distinguishable from the atmospheric neutrino flux at high energies, typically around several tens of $\mathrm{TeV}$, depending on the event selection. It is separable from the prompt atmospheric component by its energy distribution \cite{IceCube:2023mrq}.

Until recently, the astrophysical neutrino flux was consistent with a simple power-law description \cite{IceCube:2024fxo}. Several analyses using different event samples and energy ranges had revealed variations in the inferred spectral indices, though (Fig.~\ref{fig:diffuse_break}~(left)). A recent analysis now shows a clear hardening of the spectrum below about $30\,\mathrm{TeV}$ \cite{Basu:2025icrc}. Fitted with different spectrum hypotheses, the broken power-law fit yields the best agreement with the data (Fig.~\ref{fig:diffuse_break}~(right)) and is preferred over the simple power-law assumption with $4.7\sigma$. The fitted spectral indices below and above the break are $\gamma_1 = 1.7\pm0.3$ and $\gamma_2=2.8\pm0.1$, respectively.

\subsection{Flavor composition}\label{sec:flavorcomp}

\begin{figure}[t!]
\centering
\includegraphics[width=0.5\linewidth]{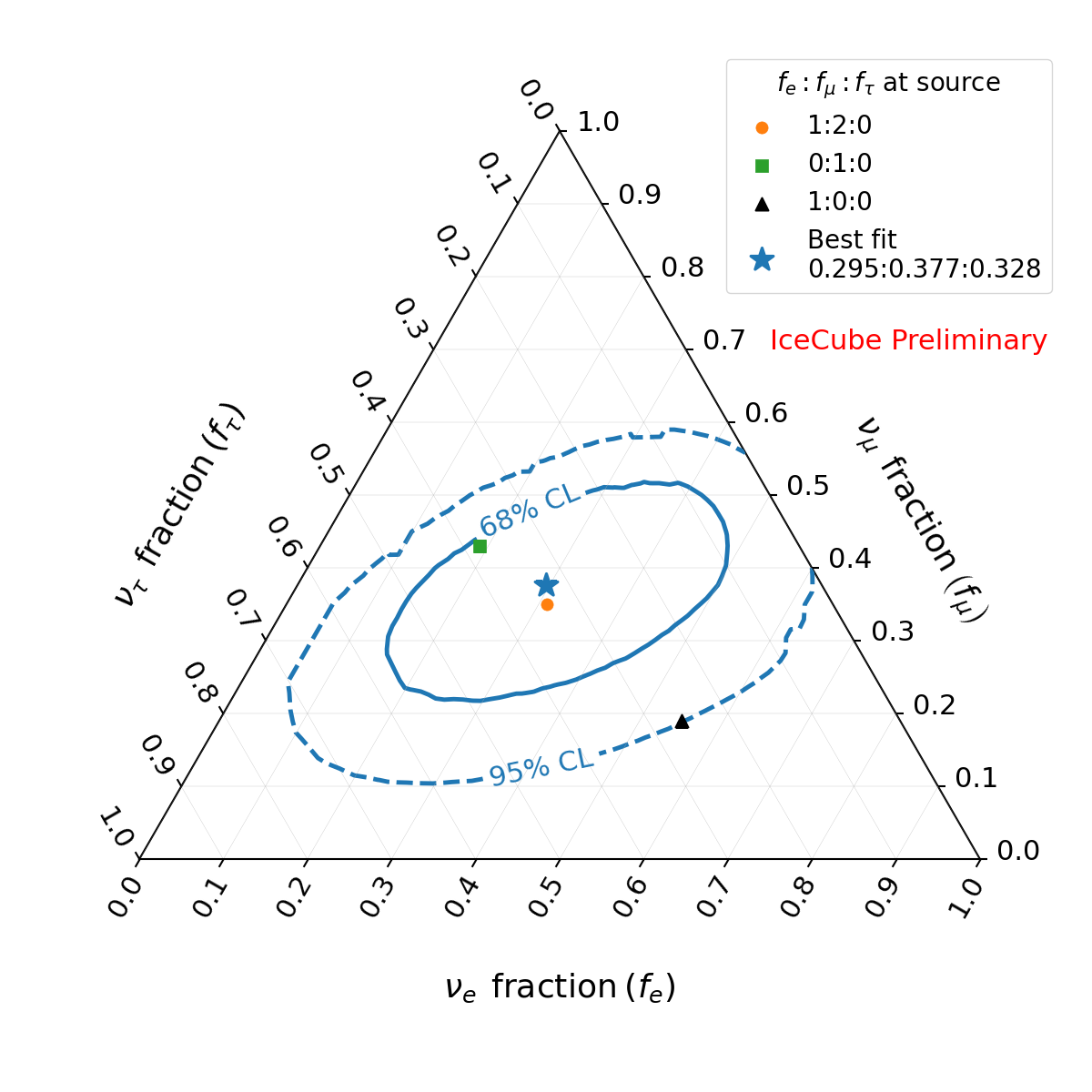}
\caption{Neutrino flavor triangle for the measured diffuse neutrino flux at Earth assuming full mixing. The best fit assumes a broken power-law. Taken from \cite{Balagopal:2025icrc}.}\label{fig:flavor}
\end{figure}

The flavor composition of an astrophysical neutrino flux carries crucial information about the production environment and the underlying acceleration mechanisms. The primary expected production channel is through pion production, in which accelerated protons or heavier nuclei interact with matter or radiation fields, generating charged pions that subsequently decay via muons into electrons and neutrinos:
\begin{align}    
    p + p/\gamma \rightarrow \pi^\pm + X \rightarrow& \mu^\pm + \overset{\scriptscriptstyle(-)}{\nu_\mu}  \\
    \mu^\pm \rightarrow& e^\pm + \overset{\scriptscriptstyle(-)}{\nu_\mu} + \overset{\scriptscriptstyle(-)}{\nu_e}
\end{align}
where $X$ represents the rest of the final state of the reaction. This results in a neutrino flavor ratio of (1:2:0) at the source. On the other hand, if the source contains strong magnetic fields, the muons loose rapidly energy so that the resulting decay neutrinos are outside the detection range of neutrino telescopes, resulting in a flavor ration of (0:1:0). A third scenario involves high energy neutrons. From their decay
\begin{equation}
    n \rightarrow p + e^- + \overset{\scriptscriptstyle(-)}{\nu_e}
\end{equation}
a flavor ratio of (1:0:0) at the source is expected.

As a consequence of neutrino oscillations, the flavor ratio at the detector is not the same as at the source, though. Since the neutrino production regions in astrophysical sources are much larger than the oscillation lengths, the oscillation term in the transition probability must be averaged. \cite{Athar:2000yw}. As a result, the expected flavor ratios at Earth for the different production scenarios are approximately: pion source (1:1:1), damped pion source (4:7:7), neutron source (5:2:2). Figure~\ref{fig:flavor} shows the current best constraints on the flavor ratio by IceCube \cite{Balagopal:2025icrc}. This analysis is based on an event sample which consists of events which have their interaction vertex contained inside the instrumented volume of IceCube, thereby allowing for an improved reconstruction of the event topologies. Assuming a broken power-law spectrum for the neutrino flux (see Sec.~\ref{sec:astro_diffuse}), for the first time, the 68\% C.L.\ contour is closed. The best fit is very well compatible with the pion production scenario and the muon-damped scenario lies at the edge of this interval. On the other hand, the neutron-source scenario can be excluded at the 95\% confidence level. It should be noted, however, that this conclusion assumes ``pure'' production scenarios, which are unlikely to occur in nature, particularly when considering the total diffuse astrophysical neutrino flux.

\subsection{EHE neutrinos and KM3-230213A}\label{sec:ehe}

\begin{figure}[t!]
\centering
\includegraphics[width=0.43\linewidth]{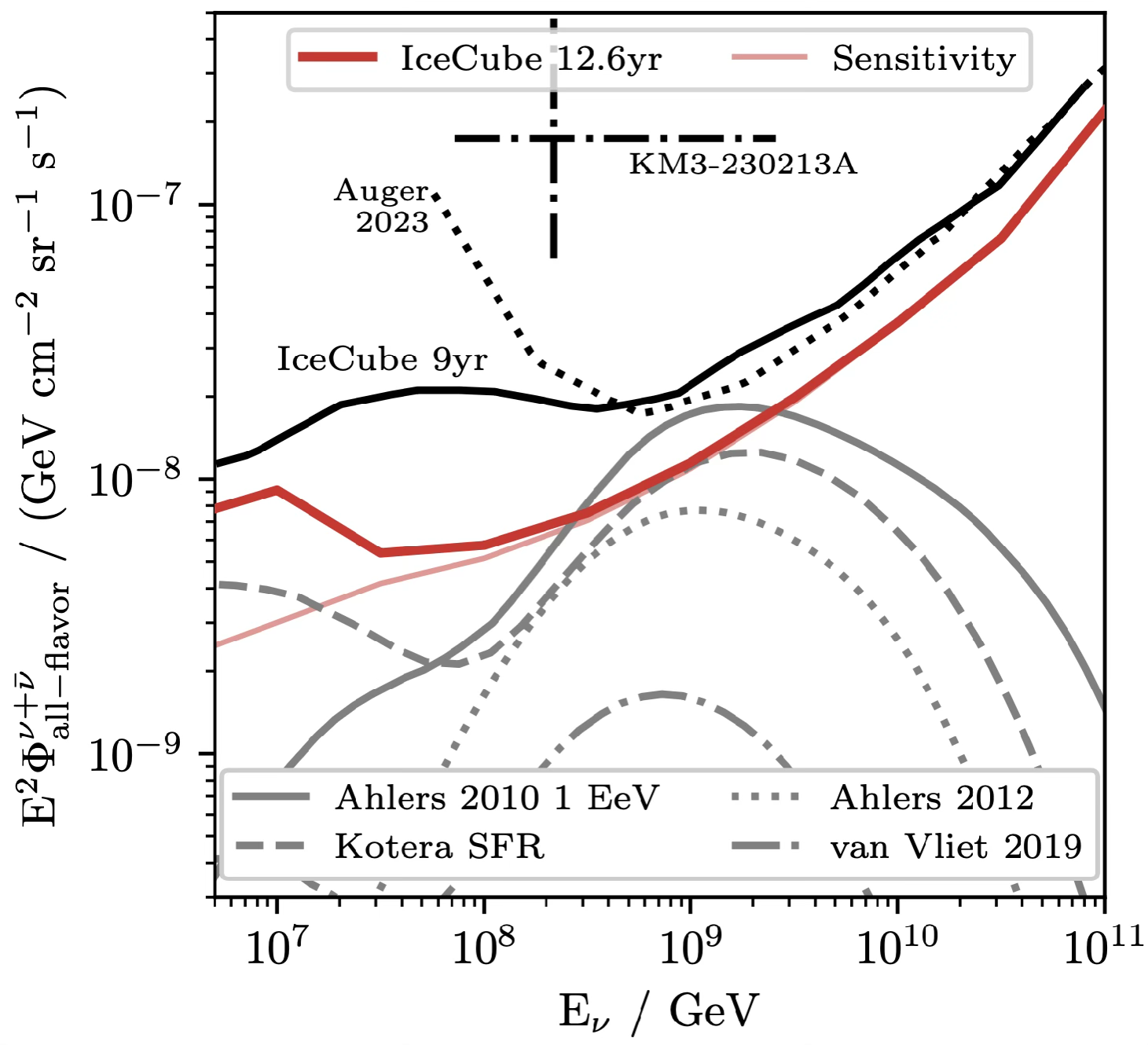}
\hfill
\includegraphics[width=0.53\linewidth]{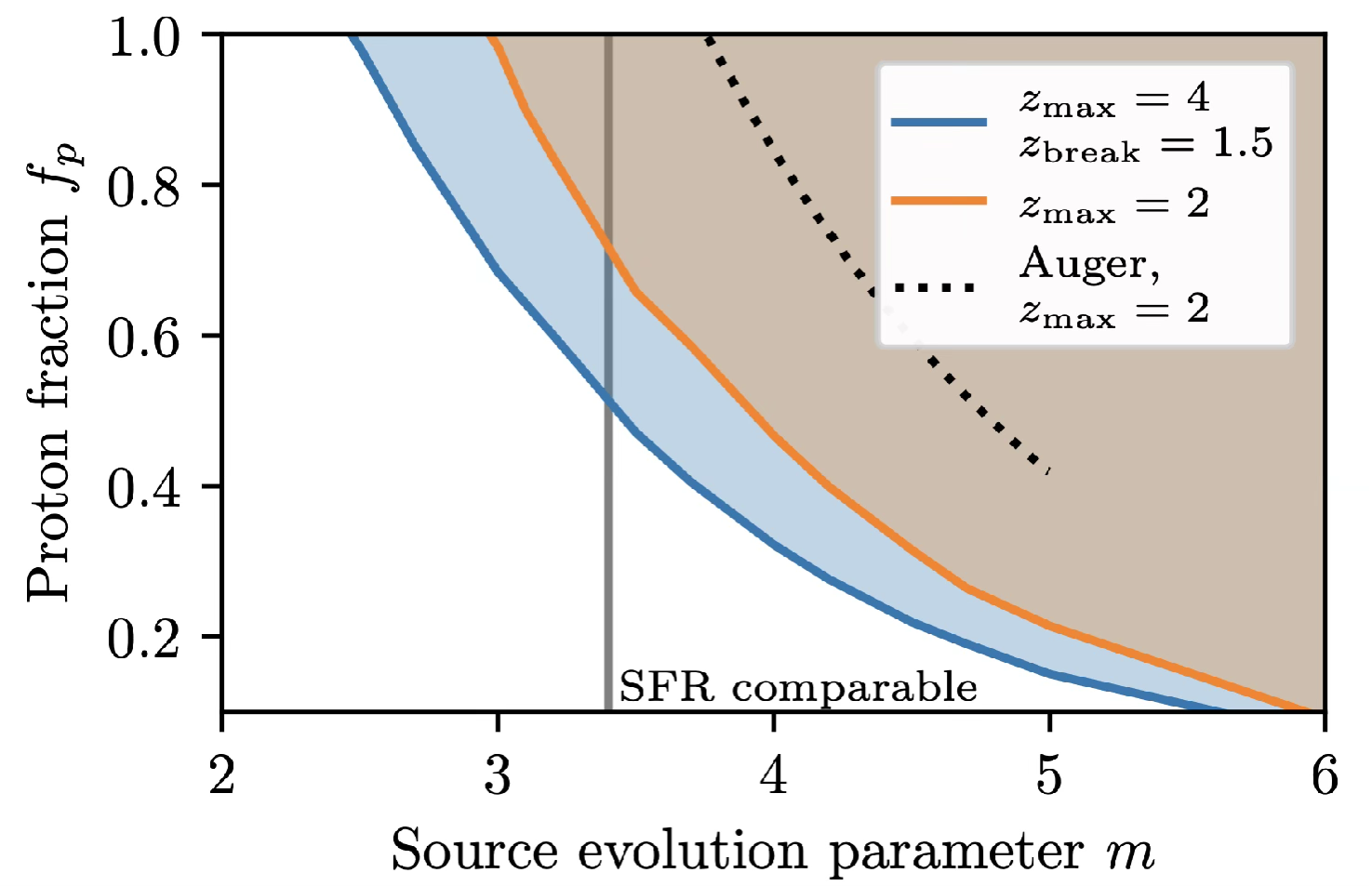}
\caption{\emph{Left:} IceCube upper limit on the cosmogenic neutrino flux as a function of energy together with the Auger upper limit and model predictions. Also shown is the flux derived from the KM3NeT event. For details and references see \cite{Meier:2025icrc}. \emph{Right:} Proton fraction of ultra-high energy cosmic rays as a function of the source evolution parameter for two source evolution models. See \cite{Meier:2025icrc} for more details. Both plots taken from \cite{Meier:2025icrc}.}\label{fig:ehe}
\end{figure}

In addition to the production scenarios described in Sec.~\ref{sec:flavorcomp}, there exists another mechanism that gives rise to so-called cosmogenic neutrinos: ultra-high energy protons with energies above $10^7\,\mathrm{GeV}$ interact with photons from the cosmic microwave background producing a neutron and a neutrino via the $\Delta$ resonance:
\begin{align}
    p + \gamma_\mathrm{CMB} &\rightarrow \Delta^+ \rightarrow n + \pi^+ \\
    \pi^+ &\rightarrow \mu^+ + \nu_\mu \\
    n &\rightarrow p + e^- + \overset{\scriptscriptstyle(-)}{\nu_e}
\end{align}
resulting in a flux of ultra-high energy neutrinos of $100\,\mathrm{PeV}$ and above. ~~\ref{fig:ehe}~(left) presents predictions from several models for this neutrino flux. IceCube has conducted searches for cosmogenic neutrinos over recent years but has not yet observed any indications of such a flux. The resulting upper limit is currently the most stringent and begins to constrain the most optimistic predictions for the cosmogenic neutrino flux. 

The fact that these neutrinos are preferentially produced by protons and much less so by heavier nuclei (for these, the threshold energy would be a factor $N$ higher where $N$ is the number of nucleons) allows to constrain the fraction of protons in the ultra-high energy cosmic rays. The limit depends on the assumed source evolution rate $(1+z)^m$ with redshift $z$ and evolution parameter $m$ (Fig.~\ref{fig:ehe}~(right)). Assuming $z_\mathrm{max} = 2$ and an evolution parameter comparable to the star formation rate results in upper limit on the proton fraction of about 0.7 .

Another notable observation in this context is the highest energy neutrino detected to date, recently reported by KM3NeT \cite{KM3NeT:2025npi} with an estimated energy of $220^{+570}_{-110}\,\mathrm{PeV}$ (68\% C.L.). When converted into a flux over the corresponding 90\% C.L.\ energy interval of $72,\mathrm{PeV}$ to $2.6,\mathrm{EeV}$, this would imply roughly 70 events in IceCube—an expectation excluded at more than $10\sigma$ by the upper limit discussed above. However, this doesn't take into account the uncertainty on the flux calculated from the KM3NeT event. Performing a joint fit of the KM3NeT event with the null observations by IceCube and Auger following \cite{KM3NeT:2025ccp}, the new upper limit by IceCube results in a moderate tension of $2.9\sigma$ with expected/observed events for KM3NeT: (0.01/1), Auger: (0.4/0) and IceCube: (0.9/0) \cite{Meier:2025icrc}. It should be noted, that these calculations are only valid if the KM3NeT event is part of a steady diffuse neutrino flux and e.g.\ not a neutrino from the burst of a single source. Due to the reduced observation time in the latter case, the limits of Auger and IceCube would be significantly weaker and the tension would be reduced considerably.

\section{Neutrinos from the Milky Way}\label{sec:milkyway}

In 2023, IceCube reported the first observation of a neutrino flux from the Milky Way at $4.5\sigma$ n\cite{IceCube:2023ame}. The corresponding analysis was based on a neutrino sample containing only cascades. An updated analysis with additional 2.5 years of data which also incorporates track events increases the sensitivity such that a detection with a significance above $5\sigma$ is expected \cite{Thiesmeyer:2025icrc}. In addition, analyses have started that fit the neutrino flux independently in different regions of the Galactic plane \cite{Neste:2025icrc,Osborn:2025icrc}. Although the currently limited statistics constrain the precision of the fitted parameters, an enhanced emission from the central region of the Milky Way is observed, as anticipated.

\section{Astrophysical neutrino sources}\label{sec:sources}

In 2022 the IceCube Collaboration reported the observation of about 80 high energy neutrinos from the nearby active galaxy NGC 1068 at a significance above $4\sigma$ \cite{IceCube:2022der}. Interestingly, no high energy gamma rays in the TeV range were observed form this source \cite{MAGIC:2019fvw}. This hints at the neutrino production in the vicinity of a supermassive black hole with high radiation density on which high-energy gamma rays are absorbed (see e.g.\ \cite{Fang:2023vdg}).

This finding prompted several follow-up analyses of IceCube data targeting neutrino emission from bright or hard X-ray sources in the northern sky, using track-like events with precise directional reconstruction \cite{IceCube:2024dou,IceCube:2024ayt}. Excluding the direction of NGC~1068, these analyses reveal hints of emission from various sources at the $\sim 3\sigma$ level. In the \emph{southern sky}, searches for neutrino emission from X-ray bright galaxies employed  tracks starting within the detector to distinguish cosmic neutrinos from the abundant atmospheric muons \cite{Yu:2025icrc}. A stacking analysis of 13 sources selected from the BASS catalog based on X-ray luminosity yields a significance of $3.0\sigma$ for neutrino emission, with Cen~A excluded to avoid contamination from jet-related emission.

Though the significances are still moderate (the results cannot be straightforwardly combined due to the presence of correlations) they support the picture of obscured AGNs as a significant source of high energy cosmic neutrinos. Further data from IceCube, KM3NeT \cite{KM3NeT_TDR_2010}, BAIKAL-GVD \cite{Malyshkin2023}, and planned future multi-km³ detectors such as IceCube-Gen2 (see Sec.~\ref{sec:icecubegen2}), P-ONE \cite{Agostini2020}, and a Chinese neutrino telescope \cite{Ye2023,Huang2025} will allow us to verify the accuracy of this picture.

\section{Dark Matter and physics beyond the Standard Model}\label{sec:dm}
Another profound mystery of the Universe is the apparent presence of an unknown form of matter, accounting for roughly 25\% of the total energy density of the Universe, inferred from a variety of observations ranging from galactic rotation curves to the cosmic microwave background \cite{Cirelli2024}. Although originally designed to detect neutrinos with energies of $1\,\mathrm{TeV}$ and above, IceCube has also proven to be a valuable instrument in the indirect search for dark matter. Dark matter in the form of Weakly Interacting Massive Particles (WIMPs) can accumulate in gravitational wells such as the Sun or the Galactic Center. Their annihilation may produce neutrinos, which could be detected by IceCube as an enhanced flux from these specific directions. Recent results can be found in \cite{Lazar:2025icrc,Chau:2025icrc,Salazar:2025icrc}. Beyond dark matter, IceCube also searches for other phenomena beyond the Standard Model, such as magnetic monopoles \cite{Häußler:2025icrc} and quantum decoherence \cite{Krishnan:2025icrc}, for which it has set world-leading limits \cite{ICECUBE:2023gdv}.

\section{Cosmic rays}\label{sec:crs}

\begin{figure}[t!]
\centering
\includegraphics[width=0.7\linewidth]{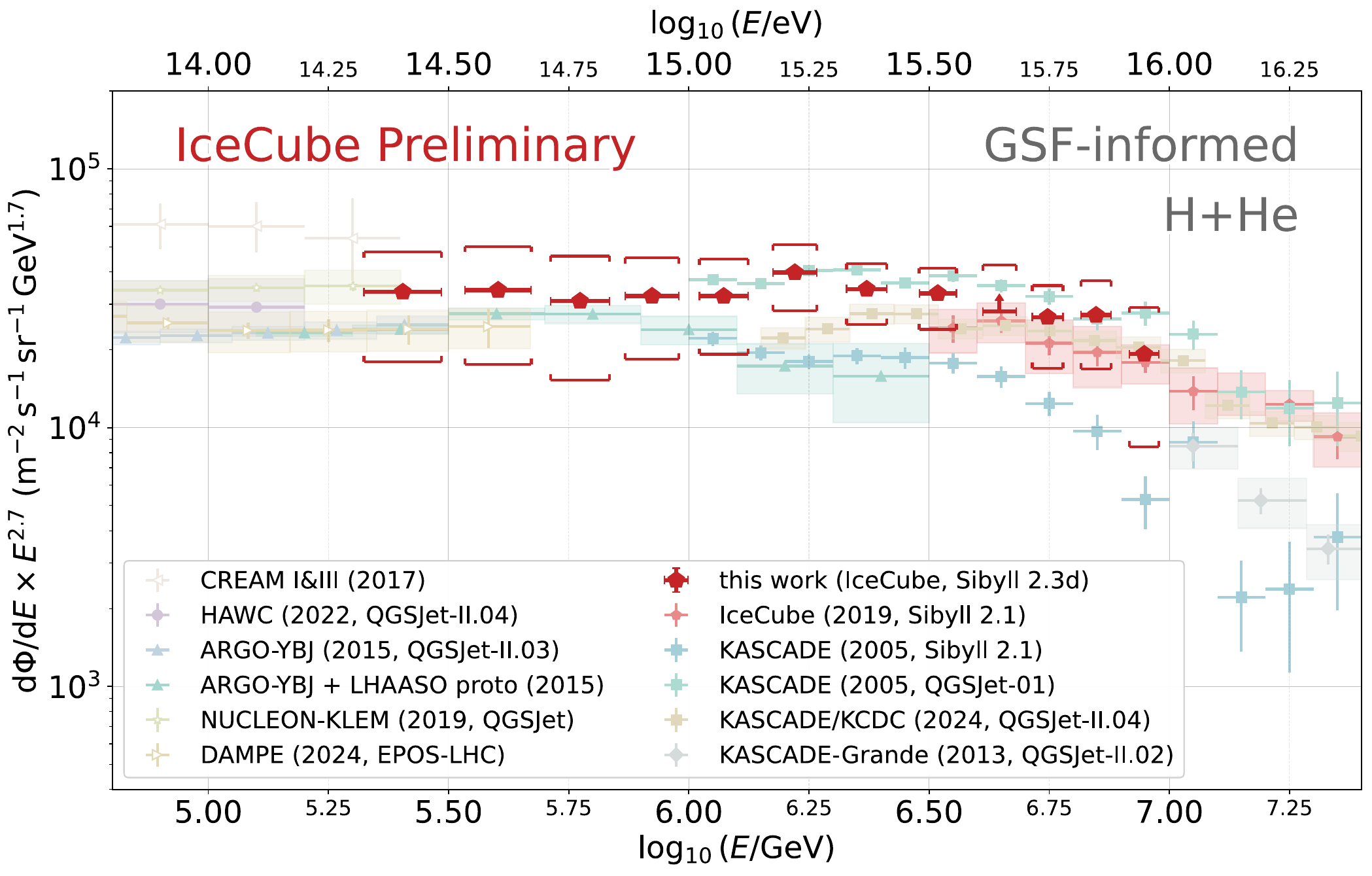}
\caption{Combined reconstructed hydrogen and helium flux. Systematic uncertainties are represented by brackets while error bars describe statistical uncertainty and energy resolution. For details refer to \cite{Saffer:2025icrc}.}\label{fig:crs_spec}
\end{figure}

In addition to the deep-ice optical array, the IceCube neutrino observatory includes the surface detector IceTop, which enables the study of the cosmic-ray energy spectrum around the knee at approximately $10^{15}\,\mathrm{eV}$ and the determination of its composition. In a recent analysis, IceTop data---sensitive to the electromagnetic and low-energy muon component of cosmic-ray air showers---were combined with measurements of TeV muons in the deep ice using a neural network approach to reconstruct the cosmic-ray energy spectrum and infer its composition \cite{Saffer:2025icrc}. This analysis was able to reconstruct the energy spectrum from $3\,\mathrm{TeV}$ down to $250\,\mathrm{TeV}$ thereby closing the gap to direct measurements (Fig.~\ref{fig:crs_spec}). It found that the spectrum in this energy region is dominated by hydrogen and helium and agrees well with that from direct measurements. Information on other analyses using IceTop can be found here \cite{Abbasi:2025icrc,Zilberman:2025icrc}. In addition, new detectors are being installed that will enhance the capabilities of the surface part of IceCube further \cite{Venugopal:2025icrc,Shefali:2025icrc,Vaidyanathan:2025icrc}.

\begin{figure}[t!]
\centering
\includegraphics[width=1\linewidth]{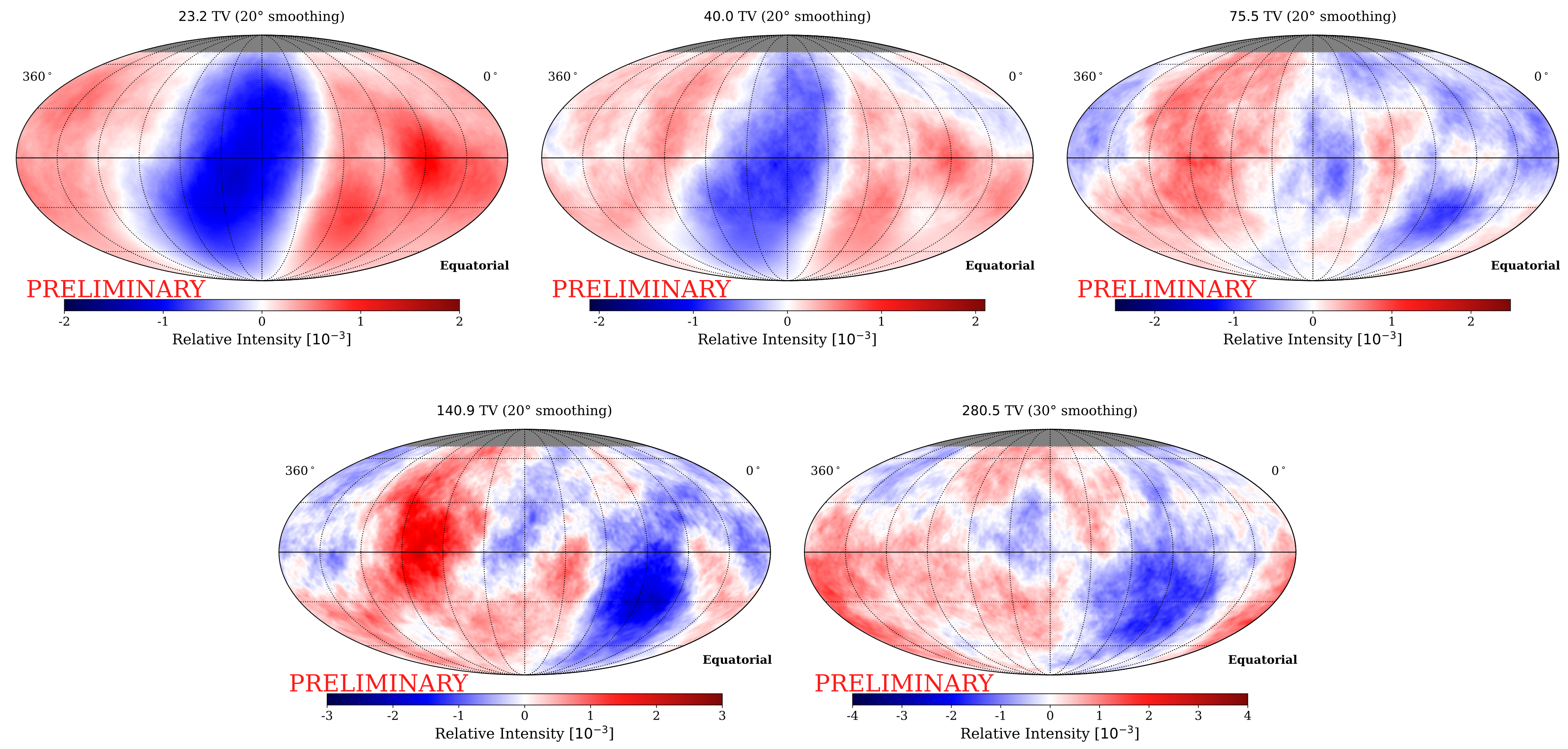}
\caption{Relative intensity for seven rigidity-matched pairs of energy bins from HAWC and IceCube. For details refer to \cite{Diaz:2025icrc}.}\label{fig:crs_anisotropy}
\end{figure}

Even on its own, the in-ice detector, recording billions of muons per year from cosmic-ray–induced air showers, provides a powerful tool for the study of cosmic rays. The huge statistics allows to search for very small anisotropies in the arrival direction of cosmic rays at Earth. As muons are absorbed in the Earth, IceCube is only sensitive to the southern sky whereas instruments like HAWC located in Mexico cover a more northern portion of the sky. Combining the analyses from these two experiments allows us to extend the coverage almost to the entire sky from $90^\circ\mathrm{S}$ to $70^\circ\mathrm{N}$ \cite{Diaz:2025icrc}. Figure~\ref{fig:crs_anisotropy} presents combined sky maps from IceCube and HAWC in seven energy bins, covering rigidities from $32.2\,\mathrm{TV}$ to $280.5\,\mathrm{TV}$. The maps show anisotropies in the intensity of the cosmic-ray flux from different directions at the $10^{-3}$ level with a rapid phase transition between $40\,\mathrm{TV}$ and $76\,\mathrm{TV}$.

\section{Into the future: IceCube Upgrade and IceCube-Gen2}\label{sec:extensions}

The recent detection of neutrinos from various cosmic sources with IceCube, together with the observation of an ultra–high-energy neutrino by KM3NeT, has highlighted the enormous potential of neutrino astronomy for probing the extreme Universe. At the same time, it has become evident that, after nearly 15 years of continuous operation of the complete IceCube detector, further data alone will yield only a moderate improvement in sensitivity to steady sources. Substantial improvements in sensitivity can instead be achieved through several approaches: 
\begin{enumerate}[label=(\alph*)]
    \item Improvement of event reconstruction and selection both in accuracy (improving the angular resolution by a factor two reduces the background for point-like sources by a factor four) and speed. The latter allows to reconstruct more events with advanced algorithms with higher angular resolution;
    \item Improvement of the detector calibration which in the case of IceCube is currently dominated by systematic uncertainties in the optical properties of the ice \cite{Chirkin:2025icrc,Dutta:2025icrc};
    \item Extension of the energy range both to higher (EeV) and lower (GeV) energies;
    \item A larger instrumented volume.
\end{enumerate}
It should be noted that items (a) and (b) can be retroactively applied to all recorded data, and several IceCube analyses have already benefited from the current Pass~2 data processing. Machine learning methods play an important role in this context \cite{Seen:2025icrc,Nakos:2025icrc,Soldin:2025icrc,Vara:2025icrc,Koundal:2025icrc}. Items (b) and (c) are presently being addressed through the low-energy extension known as the IceCube Upgrade.

\subsection{IceCube Upgrade}

\begin{figure}[t!]
\centering
\includegraphics[width=0.45\linewidth]{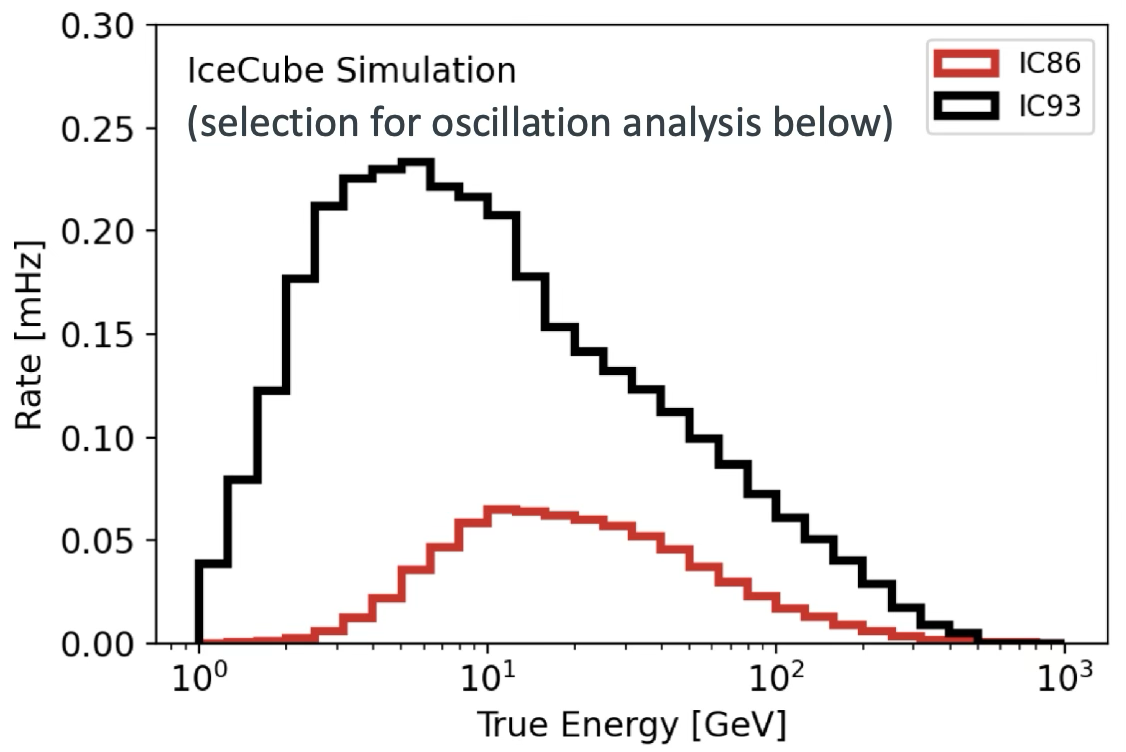}
\hfill
\includegraphics[width=0.5\linewidth]{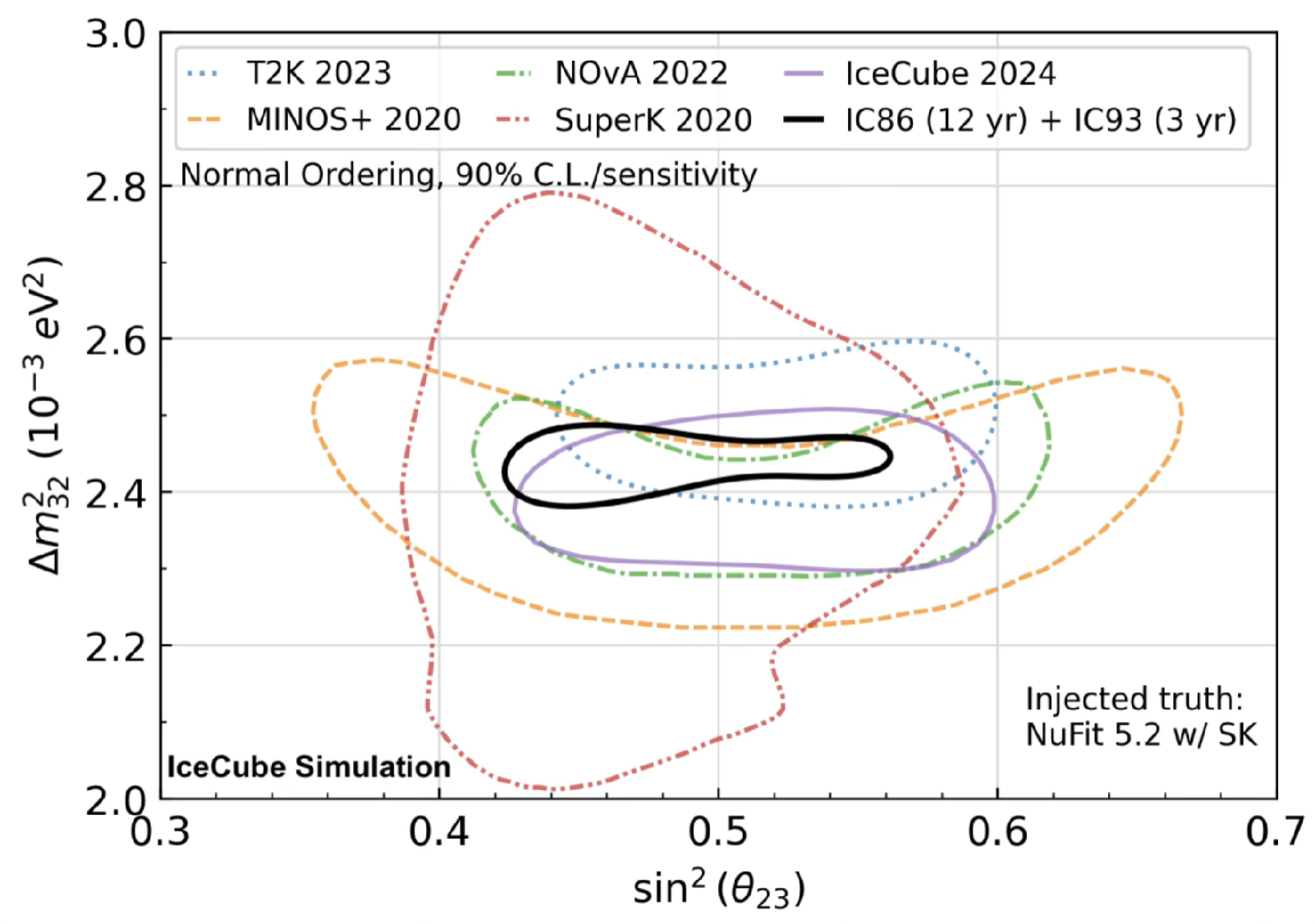}
\caption{Left: Simulated rate of detected neutrinos as a function of neutrino energy for the current IceCube detector (IC86) and the extended detector with seven additional strings from IceCube Upgrade (IC93). Right: Expected IC93 90\% confidence level sensitivity contour with 3 years of live time compared to existing measurements of  atmospheric oscillation parameters. Both figures taken from \cite{IceCube:2025chb}.}\label{fig:upgrade}
\end{figure}

IceCube Upgrade will comprise seven new strings equipped with approximately 700 advanced optical sensors and a large set of precision calibration devices. It will instrument about $2\,\mathrm{Mton}$ of clear deep ice and is scheduled for deployment during the 2025/26 austral summer at the bottom center of the existing IceCube array. Two main types of optical sensors will be used: the multi-PMT Digital Optical Module (mDOM) \cite{IceCube:2021eij,Fukami:2025icrc}, containing 24 three-inch photomultiplier tubes, and the Dual optical sensor in an Ellipsoid Glass (D-Egg) \cite{IceCube:2022mng}, equipped with two eight-inch photomultiplier tubes. With effective areas 3--4 times larger than those of the current DOMs, and significantly reduced vertical ($3\,\mathrm{m}$) and horizontal ($45\,\mathrm{m}$) spacing, the new modules will substantially increase the number of detected photons from neutrino interactions in the few-GeV range, thereby lowering IceCube’s energy threshold to approximately $1$--$2\,\mathrm{GeV}$.

Figure~\ref{fig:upgrade}~(left) shows the simulated rate of detected neutrinos as function of energy for IceCube and IceCube including the upgrade. For energies below $10\,\mathrm{GeV}$ IceCube Upgrade will increase the number of detected events by factor of about 10. This will allow for a significant improvement in various low-energy analyses including the determination of the neutrino oscillation parameters. Figure~\ref{fig:upgrade}~(right) shows the estimated sensitivity of IceCube Upgrade compared to the current best constraints on the oscillation parameters $\Delta m^2_{32}$ and $\sin^2\Theta_{23}$ by various instruments. IceCube Upgrade will allow to set world-leading constraints on these parameters and provide strictest constraints on the unitarity of the neutrino mixing matrix \cite{IceCube:2025chb}.  The extension to lower energies will also significantly broaden the accessible dark matter mass range, providing complementarity to the lower-mass region explored by direct detection experiments \cite{Lazar1:2025icrc}. Finally, together with KM3NeT/ORCA and JUNO it will have a good chance to pin down the neutrino mass ordering \cite{IceCube-Gen2:2019fet,AthaydeMarcondesdeAndre:2023vam}.

Another key objective of IceCube Upgrade is to enhance the calibration of the existing IceCube detector across its entire instrumented volume. To this end, a wide range of dedicated calibration devices will be deployed alongside the new optical sensors. Each mDOM (DEgg) features ten (12) LEDs with a wavelength of $405\,\mathrm{nm}$ close to the highest sensitivity of the optical sensors. The LEDs can inject light with known intensity and timing profiles into the ice, thereby enabling determination of the optical ice properties (scattering and absorption length) when detected by the surrounding array of photomultipliers. In addition, each sensor features three cameras which point vertically and horizontally into the ice and provide complementary information on the ice properties \cite{Rott:2025icrc,Rodan:2025icrc}. Additional calibration instruments include 22 multi-wavelength, precisely calibrated light sources (POCAM \cite{Khera:2021npv}), seven rotatable camera systems, and eleven rotatable multi-wavelength laser modules. In addition, a laser dust logger will be co-deployed attached to the end of one of the IceCube Upgrade strings and operated during deployment \cite{Eimer:2025icrc}. This will allow us to constrain the ice properties significantly better with the aim of reducing the systematic uncertainties to below $5\%$ (currently about $10\%$).

\subsection{IceCube-Gen2} \label{sec:icecubegen2}

\begin{figure}[t!]
\centering
\includegraphics[width=1\linewidth]{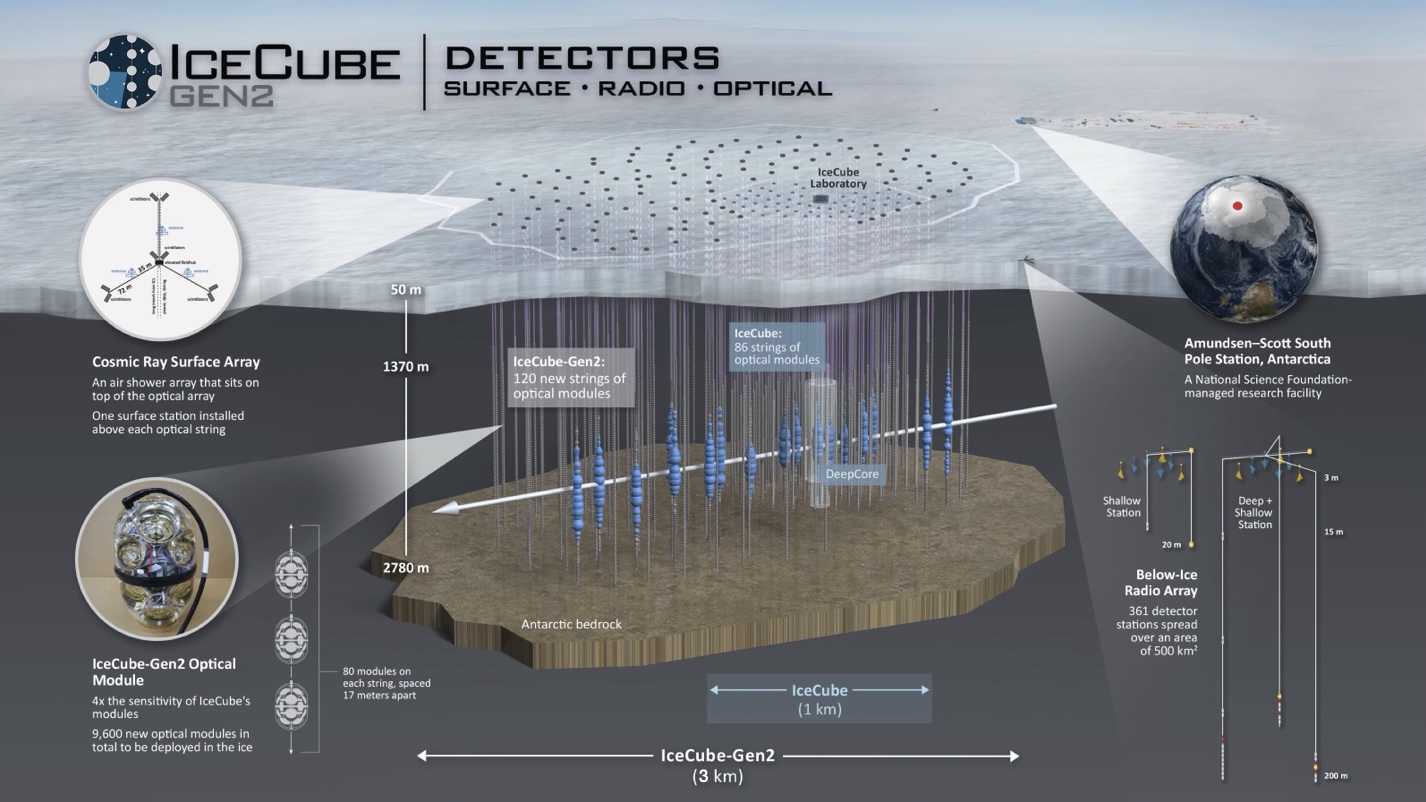}
\caption{Artist’s impression of IceCube-Gen2, highlighting several of its key components.} \label{fig:gen2_detector}
\end{figure}

\begin{figure}[t!]
\centering
\includegraphics[width=0.45\linewidth]{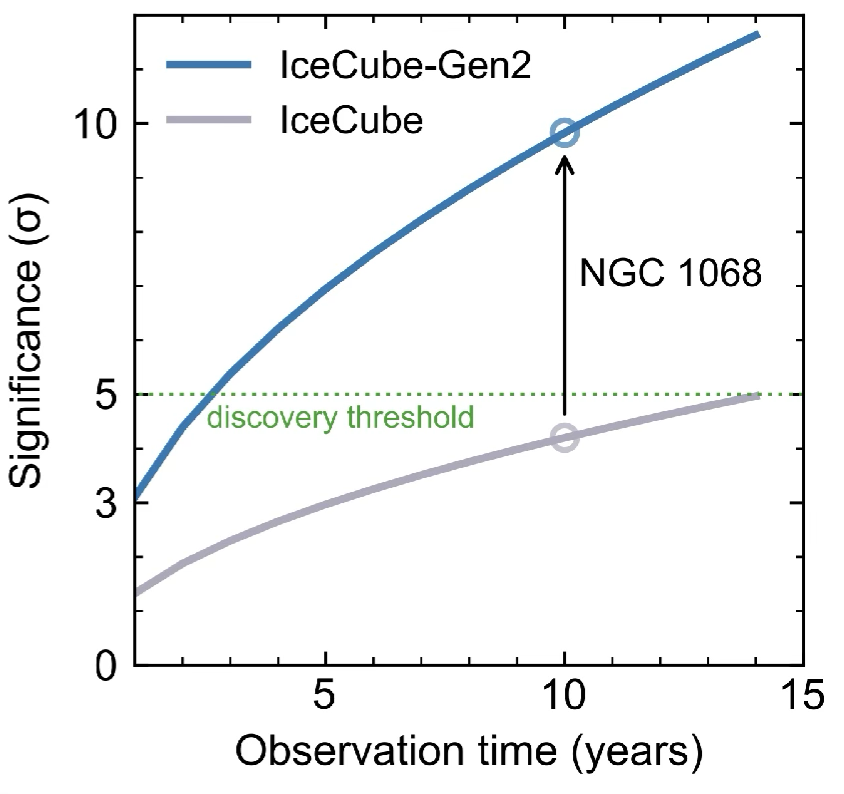}
\hfill
\includegraphics[width=0.5\linewidth]{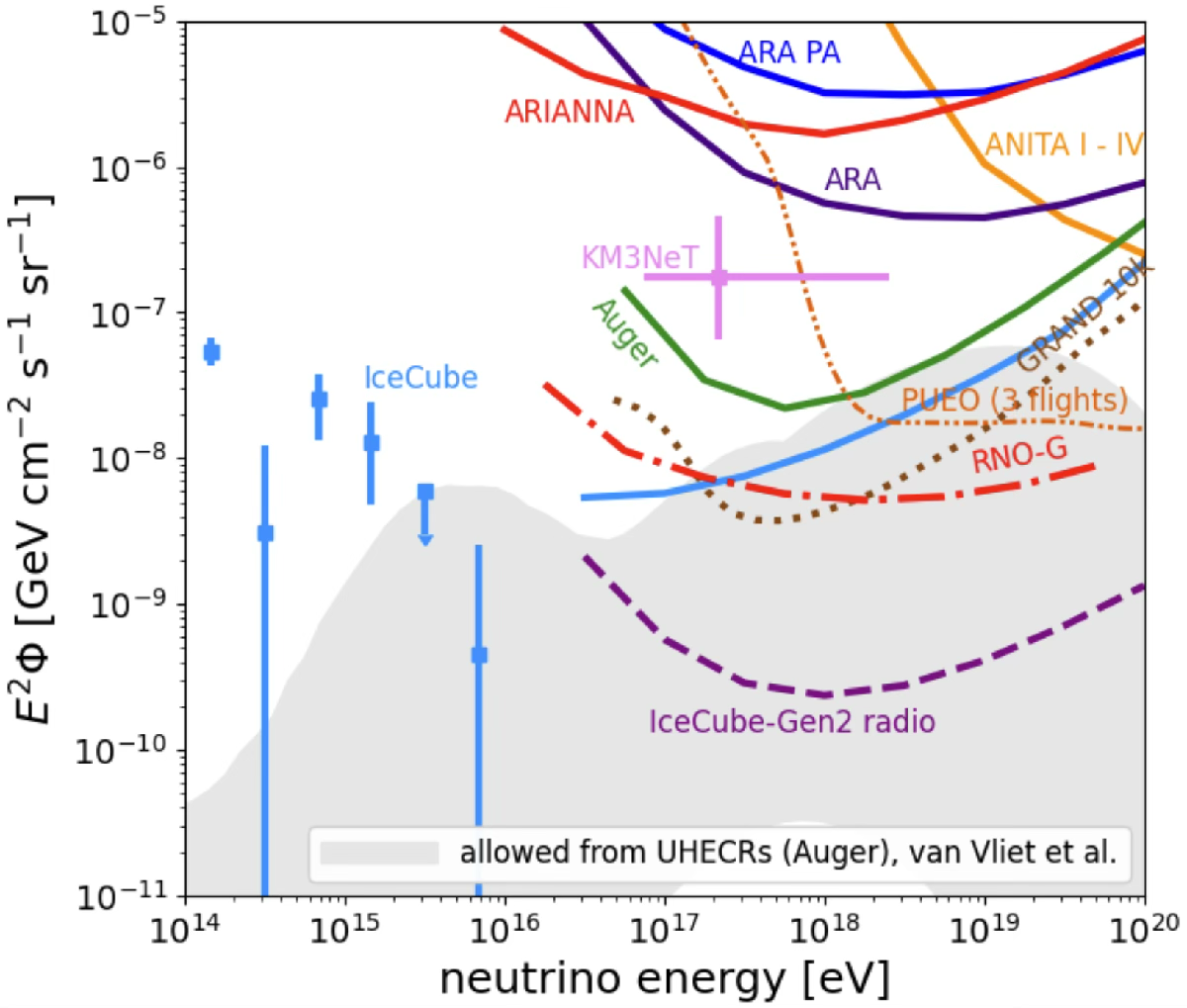}
\caption{\emph{Left:} Significance of the observed flux from NGC~1068 as a function of observation time for the current IceCube detector and IceCube-Gen2. Taken from \cite{Gen2-TDR}. \emph{Right:} Expected sensitivity of the IceCube-Gen2 in-ice radio detector at the highest energies in comparison to the allowed region of cosmogenic neutrinos based on different measurements. For details see \cite{Glaser:2025icrc}.}\label{fig:gen2-1}
\end{figure}

IceCube-Gen2 \cite{Kowalski:2025icrc} will advance neutrino astronomy at the South Pole to a new level. The planned extension includes an in-ice optical array with approximately eight times the instrumented volume of the current detector, directly addressing point (d) mentioned above. The surface detector will be extended to the footprint of the in-ice detector and consist of scintillation detectors and radio antennas \cite{Verpoest:2025icrc} for $\mathrm{PeV}$--$\mathrm{EeV}$ cosmic-ray detection \cite{Schröder:2025icrc}. Finally, a radio antenna array, sparsely deployed across approximately $500\,\mathrm{km}^2$ of ice, will provide sensitivity to ultra–high-energy neutrinos \cite{Glaser:2025icrc}. Sustainability and the environmental impact of both the construction and operation of the detector are key considerations in the planning process \cite{Khanal:2025icrc}.

With the addition of 120 new optical strings with approximately 9,600 new optical sensors, the instrumented volume of the in-ice optical detector will be increased from currently $1\,\mathrm{km}^3$ to $8\,\mathrm{km}^3$. The IceCube-Gen2 sensor will feature 16 or 18 four-inch photomultipliers housed in a pressure vessel made of borosilicate glass \cite{Kappes:2025icrc}. First prototypes were developed and will be deployed in IceCube Upgrade for in-situ testing. The expanded instrumented volume will substantially enhance IceCube’s sensitivity to PeV-scale neutrinos. Figure~\ref{fig:gen2_detector}~(left) shows the significance as a function of observation time for the neutrino flux measured from NGC~1068, comparing the current IceCube detector with IceCube-Gen2. With IceCube-Gen2 this source would have been detected at the $5\sigma$ level already after 1.5 years and with about $10\sigma$ after 10 years.

\begin{figure}[t!]
\centering
\includegraphics[width=0.8\linewidth]{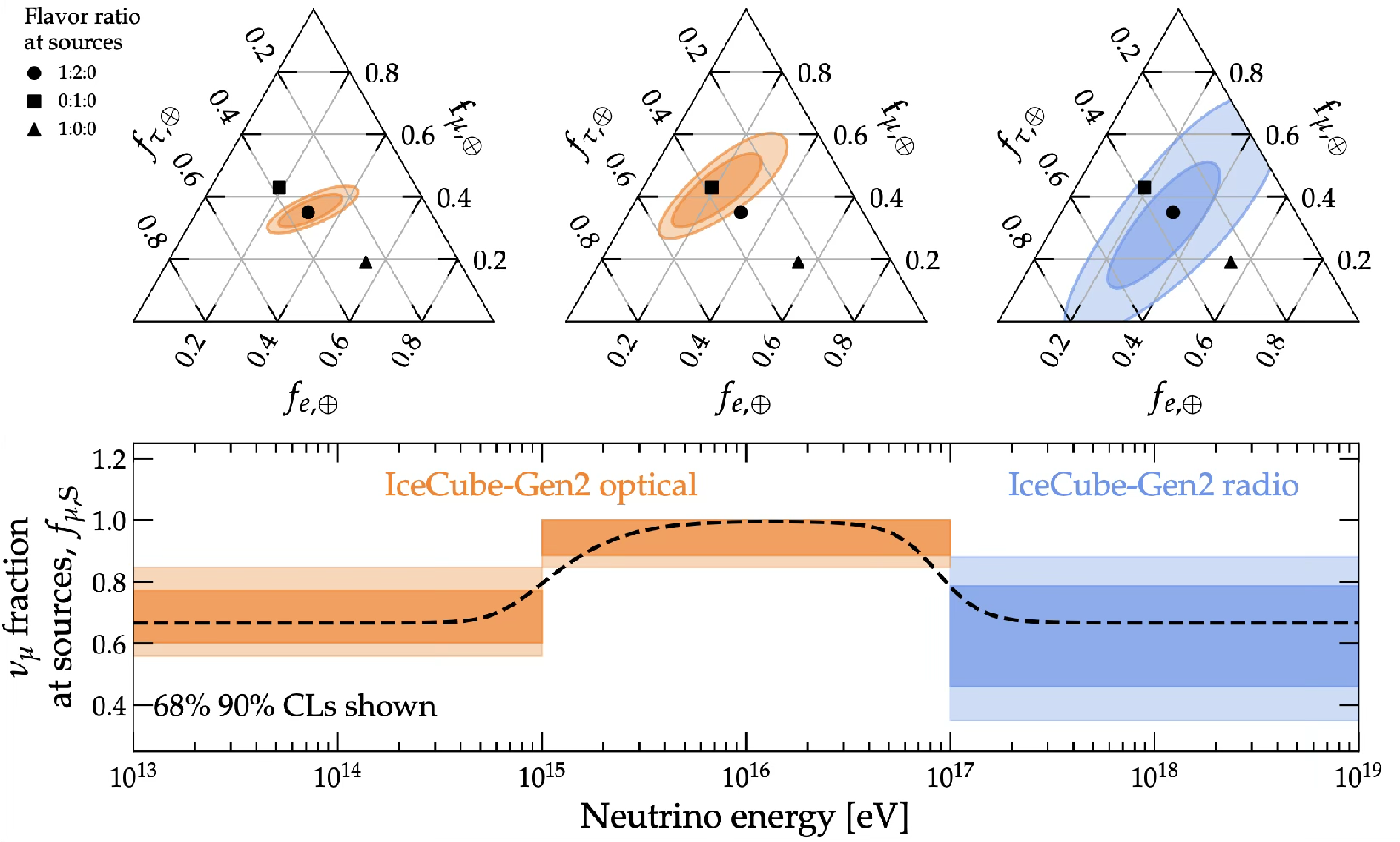}
\caption{IceCube-Gen2 capabilities to measure the neutrino flavor composition from TeV to EeV. The first two bins are measured with the optical detector of IceCube-Gen2 and the last bin with the radio
detector. Taken from \cite{Gen2-TDR}.}\label{fig:gen2_flavor}
\end{figure}

The radio array, which triggers on Askaryan emission from neutrino interactions in the ice, will allow IceCube to deeply cut into the predictions for cosmogenic neutrinos (Fig.~\ref{fig:gen2-1}~(right)). Furthermore, it will provide constraints on the population of steady and transient ultra-high energy sources above $100\,\mathrm{PeV}$ as well as enable the measurement of the neutrino flavor composition at ultra-high energies (Fig.~\ref{fig:gen2_flavor}) and neutrino-nucleon cross section.

%In summary, IceCube-Gen2 will be a powerful, unique combination of optical and radio neutrino detectors for studying atmospheric and astrophysical neutrinos from MeV to EeV energies, complemented by a powerful surface detector for studying cosmic rays. In the MeV range it will provide high statistics data on the time-dependent neutrino flux from Galactic core collapse supernovae \cite{IceCube:2023ogt,LozanoMariscal:2021the}; at GeV energies it will make world-leading contributions to the measurement of parameters in atmospheric neutrino oscillation and, via the IceCube Upgrade component, contribute to the determination of the neutrino mass hierarchy; at TeV to PeV energies, it features a ten times increased event rate for point sources and with its radio array will increase the sensitivity to neutrino flux in the EeV range by two orders of magnitude.

In summary, IceCube-Gen2 will constitute a powerful and unique combination of optical and radio neutrino detectors for the study of atmospheric and astrophysical neutrinos across the MeV--EeV energy range, complemented by a potent surface array for cosmic-ray investigations. In the MeV range, it will provide high-statistics measurements of the time-dependent neutrino flux from Galactic core-collapse supernovae \cite{IceCube:2023ogt,LozanoMariscal:2021the}; at GeV energies, it will, through the IceCube Upgrade component, provide world-leading constraints on atmospheric neutrino oscillation parameters and contribute to determining the neutrino mass ordering; at TeV–PeV energies, it will achieve an approximately tenfold increase in event rate for point sources, while the radio array will enhance sensitivity to neutrinos in the EeV range by roughly two orders of magnitude.

\section{Conclusion}\label{sec:conclusions}

The hunt for the sources of galactic and extragalactic cosmic rays is ongoing. Over the past decade, neutrinos have joined cosmic rays and electromagnetic waves---together with gravitational waves---as crucial messengers, providing key insights into the most extreme objects in the Universe. Deviations from a simple power-law spectrum in the diffuse high-energy neutrino flux have been observed, and the Galactic plane is now visible in neutrinos. Galaxies with obscured cores appear to be significant contributors to the diffuse neutrino flux, highlighting their role as potential cosmic-ray sources. Additionally, IceCube delivers world-leading measurements of atmospheric neutrino oscillations and enables direct studies of cosmic rays around the knee at $10^{15}\,\mathrm{eV}$. With the upcoming installation of IceCube Upgrade in the 2025/26 season and the planned IceCube-Gen2 in the next decade, neutrino astronomy at the South Pole is heading into an exciting future.

% Bibtex references:
\bibliographystyle{ICRC}
\setlength{\bibsep}{0pt plus 0.3ex}
\bibliography{references}

\providecommand{\href}[2]{#2}\begingroup\raggedright\begin{thebibliography}{10}

\bibitem{Mancina:2025icrc}
{\bfseries IceCube} Collaboration, S.~Mancina {\em et~al.}, {\em PoS}
  {\bfseries ICRC2025} (these proceedings) 949.

\bibitem{Raab:2025icrc}
{\bfseries IceCube} Collaboration, C.~Raab {\em et~al.}, {\em PoS} {\bfseries
  ICRC2025} (these proceedings) 956.

\bibitem{IceCube:2013low}
{\bfseries IceCube} Collaboration, M.~G. Aartsen {\em et~al.},
  \href{http://dx.doi.org/10.1126/science.1242856}{{\em Science} {\bfseries
  342} (2013) 1242856}.

\bibitem{IceCube:2018dnn}
{\bfseries IceCube, Fermi-LAT, MAGIC, AGILE, ASAS-SN, HAWC, H.E.S.S., INTEGRAL,
  Kanata, Kiso, Kapteyn, Liverpool Telescope, Subaru, Swift NuSTAR, VERITAS,
  VLA/17B-403} Collaboration, M.~G. Aartsen {\em et~al.},
  \href{http://dx.doi.org/10.1126/science.aat1378}{{\em Science} {\bfseries
  361} no.~6398, (2018) eaat1378}.

\bibitem{IceCube:2018cha}
{\bfseries IceCube} Collaboration, M.~G. Aartsen {\em et~al.},
  \href{http://dx.doi.org/10.1126/science.aat2890}{{\em Science} {\bfseries
  361} no.~6398, (2018) 147--151}.

\bibitem{IceCube:2022der}
{\bfseries IceCube} Collaboration, R.~Abbasi {\em et~al.},
  \href{http://dx.doi.org/10.1126/science.abg3395}{{\em Science} {\bfseries
  378} no.~6619, (2022) 538--543}.

\bibitem{IceCube:2023ame}
{\bfseries IceCube} Collaboration, R.~Abbasi {\em et~al.},
  \href{http://dx.doi.org/10.1126/science.adc9818}{{\em Science} {\bfseries
  380} no.~6652, (2023) adc9818}.

\bibitem{Aartsen:2016nxy}
{\bfseries IceCube} Collaboration, M.~G. Aartsen {\em et~al.},
  \href{http://dx.doi.org/10.1088/1748-0221/12/03/P03012}{{\em JINST}
  {\bfseries 12} no.~03, (2017) P03012}.

\bibitem{IceCube:2013llx}
{\bfseries IceCube} Collaboration, M.~G. Aartsen {\em et~al.},
  \href{http://dx.doi.org/10.1016/j.nima.2013.01.054}{{\em Nucl. Instrum. Meth.
  A} {\bfseries 711} (2013) 73--89}.

\bibitem{Chirkin:2025icrc}
{\bfseries IceCube} Collaboration, D.~Chirkin {\em et~al.}, {\em PoS}
  {\bfseries ICRC2025} (these proceedings) 1013.

\bibitem{IceCube:2024nhk}
{\bfseries IceCube} Collaboration, R.~Abbasi {\em et~al.},
  \href{http://dx.doi.org/10.1103/PhysRevLett.132.151001}{{\em Phys. Rev.
  Lett.} {\bfseries 132} no.~15, (2024) 151001}.

\bibitem{IceCube:2021oqo}
{\bfseries IceCube} Collaboration, R.~Abbasi {\em et~al.},
  \href{http://dx.doi.org/10.1088/1748-0221/16/08/P08034}{{\em JINST}
  {\bfseries 16} no.~08, (2021) P08034}.

\bibitem{IceCube:2013dkx}
{\bfseries IceCube} Collaboration, M.~G. Aartsen {\em et~al.},
  \href{http://dx.doi.org/10.1088/1748-0221/9/03/P03009}{{\em JINST} {\bfseries
  9} (2014) P03009}.

\bibitem{Abbasi:2021ryj}
{\bfseries IceCube} Collaboration, R.~Abbasi {\em et~al.},
  \href{http://dx.doi.org/10.1088/1748-0221/16/07/P07041}{{\em JINST}
  {\bfseries 16} (2021) P07041}.

\bibitem{IceCube:2024csv}
{\bfseries IceCube} Collaboration, R.~Abbasi {\em et~al.},
  \href{http://dx.doi.org/10.1088/1748-0221/19/06/P06026}{{\em JINST}
  {\bfseries 19} no.~06, (2024) P06026}.

\bibitem{IceCube:2024pnx}
{\bfseries IceCube} Collaboration, R.~Abbasi {\em et~al.},
  \href{http://dx.doi.org/10.3847/1538-4357/adb1de}{{\em Astrophys. J.}
  {\bfseries 981} no.~2, (2025) 182}.

\bibitem{Rootselar:2025icrc}
{\bfseries IceCube} Collaboration, L.~van Rootselaar {\em et~al.}, {\em PoS}
  {\bfseries ICRC2025} (these proceedings) 1199.

\bibitem{IceCube:2025tgp}
{\bfseries IceCube} Collaboration, R.~Abbasi {\em et~al.}, ``{Evidence for a
  Spectral Break or Curvature in the Spectrum of Astrophysical Neutrinos from 5
  TeV--10 PeV}.'' (2025).
\newblock \href{http://arxiv.org/abs/2507.22233}{{\ttfamily arXiv:2507.22233}}.

\bibitem{IceCubeCollaborationSS:2025jbi}
{\bfseries IceCube} Collaboration, R.~Abbasi {\em et~al.},
  \href{http://dx.doi.org/10.1103/PhysRevLett.135.031001}{{\em Phys. Rev.
  Lett.} {\bfseries 135} no.~3, (2025) 031001}.

\bibitem{IceCube:2024fxo}
{\bfseries IceCube} Collaboration, R.~Abbasi {\em et~al.},
  \href{http://dx.doi.org/10.1103/PhysRevD.110.022001}{{\em Phys. Rev. D}
  {\bfseries 110} no.~2, (2024) 022001}.

\bibitem{IceCube:2021rpz}
{\bfseries IceCube} Collaboration, M.~G. Aartsen {\em et~al.},
  \href{http://dx.doi.org/10.1038/s41586-021-03256-1}{{\em Nature} {\bfseries
  591} no.~7849, (2021) 220--224}. [Erratum: Nature 592, E11 (2021)].

\bibitem{Gaisser:2019efm}
T.~K. Gaisser, ``{Atmospheric Neutrinos}.'' (2019).
\newblock \href{http://arxiv.org/abs/1910.08851}{{\ttfamily arXiv:1910.08851}}.

\bibitem{Abbasi:2021qfz}
{\bfseries IceCube} Collaboration, R.~Abbasi {\em et~al.},
  \href{http://dx.doi.org/10.3847/1538-4357/ac4d29}{{\em Astrophys. J.}
  {\bfseries 928} no.~1, (2022) 50}.

\bibitem{Basu:2025icrc}
{\bfseries IceCube} Collaboration, V.~Basu {\em et~al.}, {\em PoS} {\bfseries
  ICRC2025} (these proceedings) 985.

\bibitem{IceCube:2023mrq}
{\bfseries IceCube} Collaboration, J.~Boettcher {\em et~al.},
  \href{http://dx.doi.org/10.22323/1.444.1068}{{\em PoS} {\bfseries ICRC2023}
  (2023) 1068}.

\bibitem{Balagopal:2025icrc}
{\bfseries IceCube} Collaboration, A.~Balagopal {\em et~al.}, {\em PoS}
  {\bfseries ICRC2025} (these proceedings) 983.

\bibitem{Athar:2000yw}
H.~Athar, M.~Jezabek, and O.~Yasuda,
  \href{http://dx.doi.org/10.1103/PhysRevD.62.103007}{{\em Phys. Rev. D}
  {\bfseries 62} (2000) 103007}.

\bibitem{Meier:2025icrc}
{\bfseries IceCube} Collaboration, M.~Meier {\em et~al.}, {\em PoS} {\bfseries
  ICRC2025} (these proceedings) 1122.

\bibitem{KM3NeT:2025npi}
{\bfseries KM3NeT} Collaboration, S.~Aiello {\em et~al.},
  \href{http://dx.doi.org/10.1038/s41586-024-08543-1}{{\em Nature} {\bfseries
  638} no.~8050, (2025) 376--382}. [Erratum: Nature 640, E3 (2025)].

\bibitem{KM3NeT:2025ccp}
{\bfseries KM3NeT} Collaboration, O.~Adriani {\em et~al.},
  \href{http://dx.doi.org/10.1103/yypk-zmb8}{{\em Phys. Rev. X} {\bfseries 15}
  no.~3, (2025) 031016}.

\bibitem{Thiesmeyer:2025icrc}
{\bfseries IceCube} Collaboration, M.~Thiesmeyer {\em et~al.}, {\em PoS}
  {\bfseries ICRC2025} (these proceedings) 1193.

\bibitem{Neste:2025icrc}
{\bfseries IceCube} Collaboration, L.~Neste {\em et~al.}, {\em PoS} {\bfseries
  ICRC2025} (these proceedings) 1130.

\bibitem{Osborn:2025icrc}
{\bfseries IceCube} Collaboration, J.~Osborn {\em et~al.}, {\em PoS} {\bfseries
  ICRC2025} (these proceedings) 1135.

\bibitem{MAGIC:2019fvw}
{\bfseries MAGIC} Collaboration, V.~A. Acciari {\em et~al.},
  \href{http://dx.doi.org/10.3847/1538-4357/ab3a51}{{\em Astrophys. J.}
  {\bfseries 883} (2019) 135}.

\bibitem{Fang:2023vdg}
K.~Fang, E.~L. Rodriguez, F.~Halzen, and J.~S. Gallagher,
  \href{http://dx.doi.org/10.3847/1538-4357/acee70}{{\em Astrophys. J.}
  {\bfseries 956} no.~1, (2023) 8}.

\bibitem{IceCube:2024dou}
{\bfseries IceCube} Collaboration, R.~Abbasi {\em et~al.},
  \href{http://dx.doi.org/10.3847/1538-4357/addd05}{{\em Astrophys. J.}
  {\bfseries 988} no.~1, (2025) 141}.

\bibitem{IceCube:2024ayt}
{\bfseries IceCube} Collaboration, R.~Abbasi {\em et~al.},
  \href{http://dx.doi.org/10.3847/1538-4357/ada94b}{{\em Astrophys. J.}
  {\bfseries 981} no.~2, (2025) 131}.

\bibitem{Yu:2025icrc}
{\bfseries IceCube} Collaboration, S.~Yu {\em et~al.}, {\em PoS} {\bfseries
  ICRC2025} (these proceedings) 1219.

\bibitem{KM3NeT_TDR_2010}
{\bfseries {KM3NeT}} Collaboration, ``{KM3NeT: Technical Design Report},''
  tech. rep., 2010.
\newblock
  \url{https://www.km3net.org/wp-content/uploads/2023/05/KM3NeT_DS_TDR-published-in-2010.pdf}.

\bibitem{Malyshkin2023}
Y.~Malyshkin, \href{http://dx.doi.org/10.1016/j.nima.2023.168117}{{\em NIM A}
  {\bfseries 1050} (2023) 168117}.

\bibitem{Agostini2020}
M.~Agostini and et~al., ``The pacific ocean neutrino experiment.'' (2020).
\newblock \href{http://arxiv.org/abs/arXiv:2005.09493}{{\ttfamily
  arXiv:2005.09493}}.

\bibitem{Ye2023}
Z.~P. Ye and et~al., \href{http://dx.doi.org/10.1038/s41550-023-02087-6}{{\em
  Nature Astronomy} {\bfseries 7} (2023) 1497--1505}.

\bibitem{Huang2025}
T.~Q. Huang and et~al., {\em PoS(ICRC2023)1080} (2023) .

\bibitem{Cirelli2024}
M.~Cirelli, A.~Strumia, and J.~Župan, ``Dark matter: Review of evidence,
  models, detection.'' (2024).
\newblock \href{http://arxiv.org/abs/arXiv:2406.01705}{{\ttfamily
  arXiv:2406.01705}}.

\bibitem{Lazar:2025icrc}
{\bfseries IceCube} Collaboration, J.~Lazar {\em et~al.}, {\em PoS} {\bfseries
  ICRC2025} (these proceedings) 502.

\bibitem{Chau:2025icrc}
{\bfseries IceCube} Collaboration, N.~Chau {\em et~al.}, {\em PoS} {\bfseries
  ICRC2025} (these proceedings) 476.

\bibitem{Salazar:2025icrc}
{\bfseries IceCube} Collaboration, D.~Salazar-Gallegos {\em et~al.}, {\em PoS}
  {\bfseries ICRC2025} (these proceedings) 523.

\bibitem{Häußler:2025icrc}
{\bfseries IceCube} Collaboration, J.~Häußler {\em et~al.}, {\em PoS}
  {\bfseries ICRC2025} (these proceedings) 492.

\bibitem{Krishnan:2025icrc}
{\bfseries IceCube} Collaboration, T.~Krishnan {\em et~al.}, {\em PoS}
  {\bfseries ICRC2025} (these proceedings) 1081.

\bibitem{ICECUBE:2023gdv}
{\bfseries IceCube} Collaboration, R.~Abbasi {\em et~al.},
  \href{http://dx.doi.org/10.1038/s41567-024-02436-w}{{\em Nature Phys.}
  {\bfseries 20} no.~6, (2024) 913--920}.

\bibitem{Saffer:2025icrc}
{\bfseries IceCube} Collaboration, J.~Saffer {\em et~al.}, {\em PoS} {\bfseries
  ICRC2025} (these proceedings) 376.

\bibitem{Abbasi:2025icrc}
{\bfseries IceCube} Collaboration, R.~Abbasi {\em et~al.}, {\em PoS} {\bfseries
  ICRC2025} (these proceedings) 167.

\bibitem{Zilberman:2025icrc}
{\bfseries IceCube} Collaboration, P.~Zilberman {\em et~al.}, {\em PoS}
  {\bfseries ICRC2025} (these proceedings) 458.

\bibitem{Venugopal:2025icrc}
{\bfseries IceCube} Collaboration, M.~Venugopal {\em et~al.}, {\em PoS}
  {\bfseries ICRC2025} (these proceedings) 427.

\bibitem{Shefali:2025icrc}
{\bfseries IceCube} Collaboration, Shefali {\em et~al.}, {\em PoS} {\bfseries
  ICRC2025} (these proceedings) 394.

\bibitem{Vaidyanathan:2025icrc}
{\bfseries IceCube} Collaboration, A.~Vaidyanathan {\em et~al.}, {\em PoS}
  {\bfseries ICRC2025} (these proceedings) 423.

\bibitem{Diaz:2025icrc}
{\bfseries IceCube \& HAWC} Collaboration, J.~C. Díaz~Vélez {\em et~al.},
  {\em PoS} {\bfseries ICRC2025} (these proceedings) 244.

\bibitem{Dutta:2025icrc}
{\bfseries IceCube} Collaboration, K.~Dutta {\em et~al.}, {\em PoS} {\bfseries
  ICRC2025} (these proceedings) 1029.

\bibitem{Seen:2025icrc}
{\bfseries IceCube} Collaboration, L.~Seen {\em et~al.}, {\em PoS} {\bfseries
  ICRC2025} (these proceedings) 1169.

\bibitem{Nakos:2025icrc}
{\bfseries IceCube} Collaboration, M.~Nakos {\em et~al.}, {\em PoS} {\bfseries
  ICRC2025} (these proceedings) 1127.

\bibitem{Soldin:2025icrc}
{\bfseries IceCube} Collaboration, P.~Soldin {\em et~al.}, {\em PoS} {\bfseries
  ICRC2025} (these proceedings) 1183.

\bibitem{Vara:2025icrc}
{\bfseries IceCube} Collaboration, F.~J. Vara~Carbonell {\em et~al.}, {\em PoS}
  {\bfseries ICRC2025} (these proceedings) 1201.

\bibitem{Koundal:2025icrc}
{\bfseries IceCube} Collaboration, P.~Koundal {\em et~al.}, {\em PoS}
  {\bfseries ICRC2025} (these proceedings) 1201.

\bibitem{IceCube:2025chb}
{\bfseries IceCube} Collaboration, R.~Abbasi {\em et~al.}, ``{Physics potential
  of the IceCube Upgrade for atmospheric neutrino oscillations}.'' (2025).
\newblock \href{http://arxiv.org/abs/2509.13066}{{\ttfamily arXiv:2509.13066}}.

\bibitem{IceCube:2021eij}
{\bfseries IceCube} Collaboration, R.~Abbasi {\em et~al.},
  \href{http://dx.doi.org/10.22323/1.395.1070}{{\em PoS} {\bfseries ICRC2021}
  (2021) 1070}.

\bibitem{Fukami:2025icrc}
{\bfseries IceCube} Collaboration, S.~Fukami {\em et~al.}, {\em PoS} {\bfseries
  ICRC2025} (these proceedings) 1040.

\bibitem{IceCube:2022mng}
{\bfseries IceCube} Collaboration, R.~Abbasi {\em et~al.},
  \href{http://dx.doi.org/10.1088/1748-0221/18/04/P04014}{{\em JINST}
  {\bfseries 18} no.~04, (2023) P04014}.

\bibitem{Lazar1:2025icrc}
{\bfseries IceCube} Collaboration, J.~Lazar {\em et~al.}, {\em PoS} {\bfseries
  ICRC2025} (these proceedings) 486.

\bibitem{IceCube-Gen2:2019fet}
{\bfseries IceCube-Gen2} Collaboration, M.~G. Aartsen {\em et~al.},
  \href{http://dx.doi.org/10.1103/PhysRevD.101.032006}{{\em Phys. Rev. D}
  {\bfseries 101} no.~3, (2020) 032006}.

\bibitem{AthaydeMarcondesdeAndre:2023vam}
{\bfseries KM3NeT, JUNO} Collaboration, J.~P. Athayde Marcondes~de Andr{\'e},
  N.~Chau, M.~Dracos, L.~N. Kalousis, A.~Kouchner, and V.~Van~Elewyck,
  \href{http://dx.doi.org/10.1016/j.nima.2023.168438}{{\em Nucl. Instrum. Meth.
  A} {\bfseries 1055} (2023) 168438}.

\bibitem{Rott:2025icrc}
{\bfseries IceCube} Collaboration, C.~Rott {\em et~al.}, {\em PoS} {\bfseries
  ICRC2025} (these proceedings) 1069.

\bibitem{Rodan:2025icrc}
{\bfseries IceCube} Collaboration, S.~Rodan {\em et~al.}, {\em PoS} {\bfseries
  ICRC2025} (these proceedings) 1037.

\bibitem{Khera:2021npv}
{\bfseries IceCube} Collaboration, N.~Khera and F.~Henningsen,
  \href{http://dx.doi.org/10.22323/1.395.1049}{{\em PoS} {\bfseries ICRC2021}
  (2021) 1049}.

\bibitem{Eimer:2025icrc}
{\bfseries IceCube} Collaboration, A.~Eimer {\em et~al.}, {\em PoS} {\bfseries
  ICRC2025} (these proceedings) 1034.

\bibitem{Gen2-TDR}
{\bfseries IceCube-Gen2} Collaboration, R.~Abbasi {\em et~al.}, ``{IceCube-Gen2
  Technical Design Report, Part I and II}.''
\newblock
  \url{https://icecube-gen2.wisc.edu/wp-content/uploads/2023/07/IceCube_Gen2_TDR_0.05_July27.2023_Part_I_and_II.pdf}.

\bibitem{Glaser:2025icrc}
{\bfseries IceCube} Collaboration, C.~Glaser {\em et~al.}, {\em PoS} {\bfseries
  ICRC2025} (these proceedings) 1045.

\bibitem{Kowalski:2025icrc}
{\bfseries IceCube} Collaboration, M.~Kowalski {\em et~al.}, {\em PoS}
  {\bfseries ICRC2025} (these proceedings) 1080.

\bibitem{Verpoest:2025icrc}
{\bfseries IceCube} Collaboration, S.~Verpoest {\em et~al.}, {\em PoS}
  {\bfseries ICRC2025} (these proceedings) 428.

\bibitem{Schröder:2025icrc}
{\bfseries IceCube} Collaboration, F.~Schröder {\em et~al.}, {\em PoS}
  {\bfseries ICRC2025} (these proceedings) 387.

\bibitem{Khanal:2025icrc}
{\bfseries IceCube} Collaboration, M.~Khanal {\em et~al.}, {\em PoS} {\bfseries
  ICRC2025} (these proceedings) 1159.

\bibitem{Kappes:2025icrc}
{\bfseries IceCube} Collaboration, A.~Kappes {\em et~al.}, {\em PoS} {\bfseries
  ICRC2025} (these proceedings) 1072.

\bibitem{IceCube:2023ogt}
{\bfseries IceCube} Collaboration, R.~Abbasi {\em et~al.},
  \href{http://dx.doi.org/10.3847/1538-4357/ad07d1}{{\em Astrophys. J.}
  {\bfseries 961} no.~1, (2024) 84}.

\bibitem{LozanoMariscal:2021the}
C.~J.~L. Mariscal, L.~Classen, M.~A.~U. Elorrieta, and A.~Kappes,
  \href{http://dx.doi.org/10.1140/epjc/s10052-021-09809-y}{{\em Eur. Phys. J.
  C} {\bfseries 81} no.~12, (2021) 1058}. [Erratum: Eur.Phys.J.C 82, 660
  (2022)].

\end{thebibliography}\endgroup

% Alternatively, you can include references by hand:
%\begin{thebibliography}{99}
%\bibitem{...}
%
%\end{thebibliography}

\clearpage

%The following list of authors, affiliations and funding agencies will be updated at the day of submission. The following template is a placeholder generated via https://authorlist.icecube.wisc.edu/icecube on May 17, 2025 and will be updated.
\section*{Full Author List: IceCube Collaboration}

\scriptsize
\noindent
R. Abbasi$^{16}$,
M. Ackermann$^{63}$,
J. Adams$^{17}$,
S. K. Agarwalla$^{39,\: {\rm a}}$,
J. A. Aguilar$^{10}$,
M. Ahlers$^{21}$,
J.M. Alameddine$^{22}$,
S. Ali$^{35}$,
N. M. Amin$^{43}$,
K. Andeen$^{41}$
% ...
\\
\\
% ...
$^{10}$ Universit{\'e} Libre de Bruxelles, Science Faculty CP230, B-1050 Brussels, Belgium \\
% ...
$^{16}$ Department of Physics, Loyola University Chicago, Chicago, IL 60660, USA \\
$^{17}$ Dept. of Physics and Astronomy, University of Canterbury, Private Bag 4800, Christchurch, New Zealand \\
% ...
$^{21}$ Niels Bohr Institute, University of Copenhagen, DK-2100 Copenhagen, Denmark \\
$^{22}$ Dept. of Physics, TU Dortmund University, D-44221 Dortmund, Germany \\
% ...
$^{35}$ Dept. of Physics and Astronomy, University of Kansas, Lawrence, KS 66045, USA \\
% ...
$^{39}$ Dept. of Physics and Wisconsin IceCube Particle Astrophysics Center, University of Wisconsin{\textemdash}Madison, Madison, WI 53706, USA \\
% ...
$^{41}$ Department of Physics, Marquette University, Milwaukee, WI 53201, USA \\
% ...
$^{43}$ Bartol Research Institute and Dept. of Physics and Astronomy, University of Delaware, Newark, DE 19716, USA \\
% ...
$^{63}$ Deutsches Elektronen-Synchrotron DESY, Platanenallee 6, D-15738 Zeuthen, Germany \\
$^{\rm a}$ also at Institute of Physics, Sachivalaya Marg, Sainik School Post, Bhubaneswar 751005, India \\
% ...

\subsection*{Acknowledgments}

\noindent
The authors gratefully acknowledge the support from the following agencies and institutions:
USA {\textendash} U.S. National Science Foundation-Office of Polar Programs,
U.S. National Science Foundation-Physics Division,
U.S. National Science Foundation-EPSCoR,
U.S. National Science Foundation-Office of Advanced Cyberinfrastructure,
Wisconsin Alumni Research Foundation,
Center for High Throughput Computing (CHTC) at the University of Wisconsin{\textendash}Madison,
Open Science Grid (OSG),
Partnership to Advance Throughput Computing (PATh),
Advanced Cyberinfrastructure Coordination Ecosystem: Services {\&} Support (ACCESS),
Frontera and Ranch computing project at the Texas Advanced Computing Center,
U.S. Department of Energy-National Energy Research Scientific Computing Center,
Particle astrophysics research computing center at the University of Maryland,
Institute for Cyber-Enabled Research at Michigan State University,
Astroparticle physics computational facility at Marquette University,
NVIDIA Corporation,
and Google Cloud Platform;
Belgium {\textendash} Funds for Scientific Research (FRS-FNRS and FWO),
FWO Odysseus and Big Science programmes,
and Belgian Federal Science Policy Office (Belspo);
Germany {\textendash} Bundesministerium f{\"u}r Forschung, Technologie und Raumfahrt (BMFTR),
Deutsche Forschungsgemeinschaft (DFG),
Helmholtz Alliance for Astroparticle Physics (HAP),
Initiative and Networking Fund of the Helmholtz Association,
Deutsches Elektronen Synchrotron (DESY),
and High Performance Computing cluster of the RWTH Aachen;
Sweden {\textendash} Swedish Research Council,
Swedish Polar Research Secretariat,
Swedish National Infrastructure for Computing (SNIC),
and Knut and Alice Wallenberg Foundation;
European Union {\textendash} EGI Advanced Computing for research;
Australia {\textendash} Australian Research Council;
Canada {\textendash} Natural Sciences and Engineering Research Council of Canada,
Calcul Qu{\'e}bec, Compute Ontario, Canada Foundation for Innovation, WestGrid, and Digital Research Alliance of Canada;
Denmark {\textendash} Villum Fonden, Carlsberg Foundation, and European Commission;
New Zealand {\textendash} Marsden Fund;
Japan {\textendash} Japan Society for Promotion of Science (JSPS)
and Institute for Global Prominent Research (IGPR) of Chiba University;
Korea {\textendash} National Research Foundation of Korea (NRF);
Switzerland {\textendash} Swiss National Science Foundation (SNSF).

\end{document}